\documentclass[conference]{IEEEtran}

\usepackage[english]{babel}
\usepackage{balance}
\usepackage{graphicx}
\usepackage{tikz}
\usepackage{multirow}
\usepackage{multicol}
\usepackage{booktabs}
\usepackage{url}
\usepackage{cite} 
\usepackage{enumitem}
\usepackage{caption}
\usepackage{hyperref}
\usepackage{listings}
\usepackage{mathtools}
\usepackage{amsmath}
\usepackage{xcolor}
\usepackage[normalem]{ulem}
\usepackage[english]{babel}
\usepackage[flushleft]{threeparttable}
\usepackage[all]{nowidow}
\usepackage{soul}
\usepackage{ListStyle}
\usepackage{bm}
\usepackage{tablefootnote}
\usepackage{threeparttable}
\usepackage{array, makecell}
\usepackage{colortbl}
\usepackage{amsfonts}
\usepackage{float}
\usepackage{wrapfig}

\renewcommand{\thetable}{\arabic{table}}

\newcommand*\circled[1]{\tikz[baseline=(char.base)]{
            \node[shape=circle,fill,inner sep=1pt] (char) {\textcolor{white}{#1}};}}

\definecolor{newGreen}{rgb}{0.0,0.63,.333}
\definecolor{newYellow}{rgb}{1.0,.7,.1}

\useunder{\uline}{\ulined}{}%
\DeclareUrlCommand{\bulurl}{}

\def\halfcheck{\textcolor{newYellow}{\checkmark\kern-1.1ex\raisebox{.7ex}{\rotatebox[origin=c]{125}{--}}}}
\newcommand{\greencheck}{\textcolor{newGreen}{\checkmark}}
\newcommand{\grcheck}{\textcolor{blue}{\checkmark}}

\newcommand{\xmark}{%
	\tikz[scale=0.23] {
		\draw[line width=0.7,line cap=round] (0,0) to [bend left=6] (1,1);
		\draw[line width=0.7,line cap=round] (0.2,0.95) to [bend right=3] (0.8,0.05);
}}
\newcommand{\rxmark}{\textcolor{red}{\xmark}}

\colorlet{tableheadcolor}{gray!25} 
\newcommand{\headcol}{\rowcolor{tableheadcolor}} %
\colorlet{tablerowcolor}{gray!10} 
\newcommand{\rowcol}{\rowcolor{tablerowcolor}} %
\newcommand{\topline}{\arrayrulecolor{black}\specialrule{0.1em}{\abovetopsep}{0pt}%
	\arrayrulecolor{tableheadcolor}\specialrule{\belowrulesep}{0pt}{0pt}%
	\arrayrulecolor{black}}
\newcommand{\midline}{\arrayrulecolor{tableheadcolor}\specialrule{\aboverulesep}{0pt}{0pt}%
	\arrayrulecolor{black}\specialrule{\lightrulewidth}{0pt}{0pt}%
	\arrayrulecolor{white}\specialrule{\belowrulesep}{0pt}{0pt}%
	\arrayrulecolor{black}}

\newcommand{\bottomline}{\arrayrulecolor{white}\specialrule{\aboverulesep}{0pt}{0pt}%
	\arrayrulecolor{black}\specialrule{\heavyrulewidth}{0pt}{\belowbottomsep}}%
\newcommand{\bottomlinec}{\arrayrulecolor{tablerowcolor}\specialrule{\aboverulesep}{0pt}{0pt}%
	\arrayrulecolor{black}\specialrule{\heavyrulewidth}{0pt}{\belowbottomsep}}%

\begin{document}

\title{\huge{Resolving Indirect Calls in Binary Code via Cross-Reference Augmented Graph Neural Networks}}

\author{
    \IEEEauthorblockN{
        Haotian Zhang\textsuperscript{*}\IEEEauthorrefmark{2},
        Kun Liu\textsuperscript{*}\IEEEauthorrefmark{3},
        Cristian Garces\IEEEauthorrefmark{3},
        Chenke Luo\IEEEauthorrefmark{3},
        Yu Lei\IEEEauthorrefmark{4}, and
        Jiang Ming\IEEEauthorrefmark{3}
    }
    \IEEEauthorblockA{
        \IEEEauthorrefmark{2}New Jersey Institute of Technology,
        \IEEEauthorrefmark{3}Tulane University,
        \IEEEauthorrefmark{4}University of Texas at Arlington
    }
    \IEEEauthorblockA{
        \{kliu14, cgarces1, cluo6, jming\}@tulane.edu,
        haotian.zhang@mavs.uta.edu,
        ylei@cse.uta.edu
    }
}


\maketitle

\begingroup\def\thefootnote{*}\footnotetext{These authors contributed equally to this work.}\endgroup

\pagestyle{plain}
\newcommand{\mytool}{CupidCall}

\begin{abstract}



Binary code analysis is essential in scenarios where source code is unavailable, with extensive applications across various security domains. However, accurately resolving indirect call targets remains a longstanding challenge in maintaining the integrity of static analysis in binary code. This difficulty arises because the operand of a call instruction (e.g., \texttt{call rax}) remains unknown until runtime, resulting in an incomplete inter-procedural control flow graph (CFG). Previous approaches have struggled with low accuracy and limited scalability. To address these limitations, recent work has increasingly turned to machine learning (ML) to enhance analysis.  However, this ML-driven approach faces two significant obstacles: low-quality callsite-callee training pairs and inadequate binary code representation, both of which undermine the accuracy of ML models.



In this paper, we introduce \emph{\mytool}, a novel approach for resolving indirect calls using graph neural networks. Existing ML models in this area often overlook key elements such as data and code cross-references, which are essential for understanding a program's control flow. In contrast, \mytool augments CFGs with cross-references, preserving rich semantic information. Additionally, we leverage advanced compiler-level type analysis to generate high-quality callsite-callee training pairs, enhancing model precision and reliability. We further design a graph neural model that leverages augmented CFGs and relational graph convolutions for accurate target prediction.
Evaluated against real-world binaries from GitHub and the Arch User Repository on x86\_64 architecture, \mytool achieves an F1 score of 95.2\%, outperforming state-of-the-art ML-based approaches. These results highlight \mytool’s effectiveness in building precise inter-procedural CFGs and its potential to advance downstream binary analysis and security applications.

\end{abstract} 	



\section{Introduction}

Indirect call (icall) target prediction poses a significant challenge in binary code analysis tasks as it is inherently undecidable, presenting a fundamental obstacle in constructing complete inter-procedural control flow graphs (CFGs). 
The dynamic nature of icalls complicates static analysis, as targets of these calls are not explicitly known until runtime.
Mitigating this challenge would provide deeper insights into a program's behavior, 
thereby advancing capabilities in various binary analysis domains, such as binary rewriting~\cite{Gregory20,Sushant20,Meng21}, recompilation~\cite{Hasabnis16,BinRec20,Egalito}, and software security~\cite{Kuznetzov14,miTrimmer,Priyadarshan24}.

Dynamic analysis can be employed to address the challenge of indirect call resolution but faces significant limitations due to inadequate code coverage. 
One of the primary issues with dynamic analysis is its reliance on test suites, which often fail to provide comprehensive coverage. 
Techniques such as fuzzing and concolic execution are commonly used in dynamic analysis to improve coverage. For example, fuzzing tools~\cite{zalewski2017afl,Kostya2015libFuzzer} 
generate random or semi-random inputs for a program to discover new execution paths. Despite their effectiveness in uncovering hidden paths, fuzzing often fails to reach paths with deep and conditional logic, 
as they heavily depend on specific input patterns~\cite{Metzman2021FuzzBench}.
Concolic execution~\cite{KLEE,CUTE}, which combines concrete and symbolic execution, addresses more complex logic by utilizing symbolic values alongside concrete data to systematically explore feasible execution paths. 
While more thorough, concolic execution can be hampered by state space explosion, rendering it an impractical solution for large-scale applications. 
These limitations make dynamic analysis an unreliable method for collecting comprehensive data on icall targets, which is crucial for constructing robust machine learning datasets.


Static binary analysis, while capable of examining and resolving indirect calls without executing the program, 
struggles with the absence of type information in stripped binaries, which impacts the accuracy of icall resolution~\cite{Meng16}. 
Techniques like value set analysis (VSA)~\cite{Guo2019DeepVSA, Balakrishnan10, Redini19BinTrimmer} approximate all possible values a variable can assume, aiding in predicting potential icall targets by analyzing pointer value ranges. 
However, VSA can be conservative and imprecise, leading to over-approximations that include many impossible targets, resulting in false positives and reduced effectiveness~\cite{Balakrishnan10, Baldoni18}.
Another approach is symbolic execution~\cite{King1976OG_symbolicExecution,angr2016artofwar}, which finds icall targets by solving symbolic equations for path conditions. Nonetheless, symbolic execution is computationally expensive due to the state explosion problem~\cite{angr2016artofwar}, where the number of symbolic paths grows exponentially with program size, limiting its scalability in large-scale binary analysis.

With the shift to applying machine learning (ML) in multiple binary code analysis tasks~\cite{debin,pei2020trex,DeepBinDiff,pei2021stateformer,jin2022symlm,david2020neural,he2024code}, improvements have been observed in icall resolution~\cite{Callee, AttnCall2024}. However, most ML models still face performance and generalizability issues due to their icall target collection mechanisms, binary code representation, and tokenization procedures. These issues partly arise from the use of traditional static or dynamic analysis tools to collect callsite-callee pairs, 
which introduces inaccuracies due to the previously discussed weaknesses.
A key limitation of current ML models for binary analysis lies in the common practice of representing assembly code as a natural language sequence. This approach incorrectly assumes linear relationships within assembly code, similar to those found in natural languages. In reality, assembly code exhibits complex control flow and data dependencies that are not captured by linear sequences. Consequently, existing ML models often suffer from information loss as they fail to represent these complex code structures and hidden relationships, ultimately hindering their performance~\cite{debin,jin2022symlm,he2024code}.

\begin{figure} [t]
	\centering
	\includegraphics[width=0.99\columnwidth]{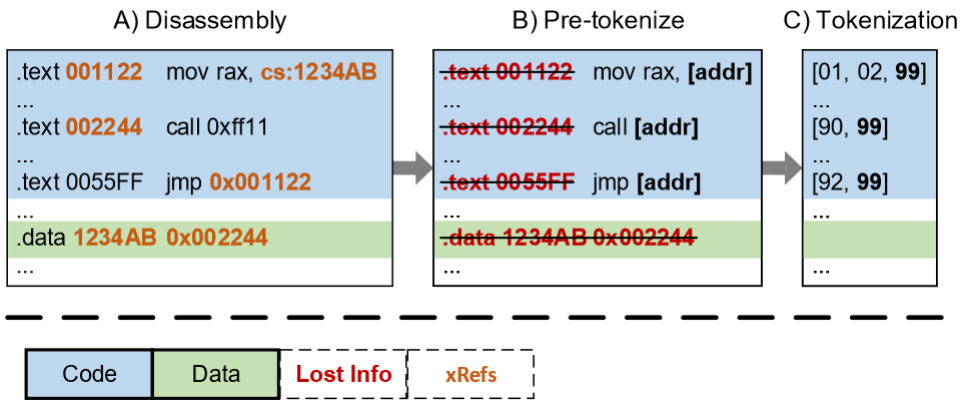}
	\vspace{-2mm}
	\caption{An example of information loss. \textbf{A)} This example shows three types of address xRefs in code and data sections: code-to-code (jmp instruction), code-to-data (mov instruction), and data-to-code (data section, which is a function pointer).  \textbf{B)} Replacing numeric values such as data and addresses with special tokens is common in traditional NLP tokenization processes to avoid out-of-vocabulary issues. \textbf{C)} After tokenization, 
 it becomes impossible to determine which \textbf{[addr]} tokens were cross-referenced.}
	\label{fig:infoloss}
	\vspace{-5mm}
\end{figure}

To illustrate the information lost during instruction encoding, consider Figure~\ref{fig:infoloss}. The standard preprocessing prodedure involves removing address and data information and replacing addresses with a dummy token, such as ``[addr]''. This practice is traditionally employed to mitigate out-of-vocabulary issues. However, as a result, the model becomes incapable of learning the hidden relationships, as data and code/data cross-references (xRefs) are lost.
Although AI-assisted icall target prediction has demonstrated superior performance over traditional heuristics-based methods~\cite{TypeArmor,Lin2023TypeSqueezer,BPA}, we hypothesize that incorporating such missing information into AI models can significantly  enhance their predictive accuracy and overall performance.

To this end, we propose a novel approach, called \emph{\mytool}, to predict icall targets in stripped binaries using graph neural networks (GNNs). 
\mytool employs symbolization~\cite{angr2016artofwar, pang2021sok} to explore xRefs information in binaries and integrating them into CFGs. 
These augmented CFGs, along with the high-quality callsite-callee pairs we collected, are used to enhance the training of heterogeneous GNNs.
A key advantage of \mytool is its use of an advanced compile-level control flow integrity (CFI) technique, TyPro~\cite{bauer2022typro}, to collect icallsite-callee pairs as training data. 
TyPro performs type propagation to refine icall targets, avoiding path explosion by focusing type propagation only on code where the type is created, used, and modified~\cite{bauer2022typro}. 
TyPro's access to rich type information allows it to perform this task with high precision.
Furthermore, we customize TyPro and the linker so that TyPro's results can be effectively recognized by our AI model.

We performed a set of experiments to evaluate \mytool using real-world projects collected from GitHub and the Arch User Repository, resulting in a corpus of $2,680$ stripped binary files compiled with O0$\sim$O3 optimization levels on x86\_64 architecture. 
Our results demonstrate \mytool's superior performance compared to the current state-of-the-art model, Callee~\cite{Callee}, in identifying indirect call targets. 
Specifically, \mytool has a F1 score of 95.2\% compared to Callee's 89.9\%.
Further analysis of \mytool's improvements against Callee indicates additional advancements, 
particularly with respect to minimizing precision degradation in call site matching, with precision and recall rates of 97.1\% and 93.3\%, respectively. 

In a nutshell, we make the following key contributions:

\begin{itemize}
	\item We propose a GNN-based model to accurately predict indirect call targets in stripped binaries. \mytool excels in minimizing precision degradation, thereby enhancing various downstream applications.
    \item We adapt heterogeneous GNNs to operate on CFGs augmented with xRef edges and data nodes, a representation that preserves richer semantic links compared to prior linear or homogeneous graph encodings., resulting in improved representation learning. This enhancement is expected to further advance other GNN-based binary analysis tasks. 
	\item We offer high-quality icall training datasets derived from compiler-level type analysis. To the best of our knowledge, this is the largest collection of icallsite-callee pairs so far, facilitating future research.

	
\end{itemize}

\noindent \textbf{Open Source }
We release a prototype of \mytool, the pre-trained model, and datasets
to facilitate reproduction, replication, and reuse, as all are found at \href{https://zenodo.org/records/12364897}{\underline{Zenodo}}.

\section{Background and Related Work}\label{sec:background2}
We first provide background information to explain why icall resolution is important. 
Then, we review existing works that attempt to tackle the icall resolution challenge and identify their pitfalls. 
Lastly, we introduce the key technical components that we leveraged to perform this study.

\newcolumntype{P}[1]{>{\centering\arraybackslash}p{#1}}

\begin{table*}[t]
	\centering
    \caption{Comparison of representative approaches on resolving indirect call targets at the binary level.}
    \label{table:CurrentApproaches}
    \vspace{-1mm} 
    \resizebox{0.88\textwidth}{!}{
	\begin{threeparttable}
		\begin{tabular}{lccccr}
			
			\topline
			\headcol 
			%
			 & \textbf{How to Obtain Training Data} & \textbf{Key Methodology} & \thead{\textbf{Considers} \\ \textbf{data section?}} & \thead{\textbf{Considers} \\ \textbf{xRef info?}}  &\textbf{F1 Score} \\ 
			\raisebox{-0.5ex}[-0.5ex]{\mytool (our work)} &  \raisebox{-0.5ex}[-0.5ex]{Compiler-level Type Analysis}  & \raisebox{-0.5ex}[-0.5ex]{BERT + GNN} & \raisebox{-0.5ex}[-0.5ex]{$\pmb{\greencheck}$} & \raisebox{-0.5ex}[-0.5ex]{$\pmb{\greencheck}$} & \raisebox{-0.5ex}[-0.5ex]{95.2}\%\\  

			\rowcol
			\raisebox{-2ex}[-2ex]{Callee~\cite{Callee}} & \raisebox{-2ex}[-2ex]{\thead{Hardware Tracing + \\ Emulation}} & \raisebox{-2ex}[-2ex]{doc2vec + DNN}  & \raisebox{-2ex}[-2ex]{\rxmark} & \raisebox{-2ex}[-2ex]{\rxmark} & \raisebox{-2ex}[-2ex]{89.9\%\tnote{a}} \\ 
			
			\raisebox{-2ex}[-2ex]{AttnCall~\cite{AttnCall2024}} & \raisebox{-2ex}[-2ex]{\thead{Callsites and Callees \\ of Direct Calls}}  & \raisebox{-2ex}[-2ex]{Transformer + DNN} & \raisebox{-2ex}[-2ex]{\rxmark} & \raisebox{-2ex}[-2ex]{\rxmark} & \raisebox{-2ex}[-2ex]{N/A\tnote{b}}
			\\ 
			
			\rowcol
			\raisebox{-0.4ex}[-0.4ex]{TypeArmor~\cite{TypeArmor}}  & \raisebox{-0.4ex}[-0.4ex]{N/A} &  \thead{Static Type-based Matching} 
			 & \raisebox{-0.4ex}[-0.4ex]{$\pmb{\halfcheck}$} & \raisebox{-0.4ex}[-0.4ex]{$\pmb{\halfcheck}$} & 
			 \raisebox{-0.4ex}[-0.4ex]{51.9\%} \\ 
			
			\raisebox{-0.4ex}[-0.4ex]{TypeSqueezer~\cite{Lin2023TypeSqueezer}}  & \raisebox{-0.4ex}[-0.4ex]{N/A} & \thead{Dynamic Type Inference +  \\ Static Type-based Matching} & 
			\raisebox{-0.4ex}[-0.4ex]{$\pmb{\halfcheck}$} & \raisebox{-0.4ex}[-0.4ex]{$\pmb{\halfcheck}$ } & 
			\raisebox{-0.4ex}[-0.4ex]{N/A \tnote{c}} \\ 
			\rowcol
			\raisebox{-0.5ex}[-0.5ex]{BPA~\cite{BPA}} & \raisebox{-0.5ex}[-0.5ex]{N/A} & \raisebox{-0.5ex}[-0.5ex]{Value Set Analysis (VSA)} & \raisebox{-0.5ex}[-0.5ex]{ $\pmb{\halfcheck}$ } & 
			\raisebox{-0.5ex}[-0.5ex]{$\pmb{\halfcheck}$} &\raisebox{-0.5ex}[-0.5ex]{73.1\%} \\ 
			
			\bottomlinec
			
		\end{tabular}
		\begin{tablenotes}
			
			\footnotesize
			\item[a] Callee's score is reported after our optimization. Callee's un-optimized F1 score is 43.9\%. For more reproduction details, please refer to $\S$\ref{sec:CalleeComparison}.
			\item[b] AttnCall is not reproducible and their reported score is derived from training and testing only on direct calls. 
			\item[c] TypeSqueezer does not report accuracy metrics and only provides the average indirect call targets (AICT) code discovery metrics. 
		\end{tablenotes}
	\end{threeparttable}    
 }
\vspace{-5mm}
\end{table*}

Due to their inherently ambiguous and imprecise nature, resolving icall targets poses a significant challenge for various binary analysis tasks,
such as binary disassembly~\cite{Priyadarshan24,pang2021sok}, rewriting~\cite{Gregory20,Sushant20,Meng21}, debloating~\cite{miTrimmer,Redini19BinTrimmer,Nibbler}, and recompilation~\cite{Hasabnis16,BinRec20,Egalito}. 
Unlike direct calls, where the call target is explicitly provided, the target of an icall is not determined until runtime. 
In cyberattacks, this reliance on runtime information allows malicious parties to exploit icalls for nefarious purposes. Specifically, attackers can leverage icalls to perform code-reuse attacks (CRA) by manipulating the program 
to utilize existing code segments within the binary~\cite{Shacham2007,Roemer12}. 
A mainstream countermeasure against CRA involves protecting indirect transfers from going to unintended locations, commonly referred to as control flow integrity (CFI)~\cite{abadi2005control-flow}.

%





\subsection{Traditional Indirect Call Resolution}

Traditional methods on icall target resolution~\cite{TypeArmor,Lin2023TypeSqueezer,BPA}, limited by the absence of type information in stripped binaries,
suffer from low accuracy or poor scalability.  
TypeArmor~\cite{TypeArmor} uses relaxed rules such as argument count bounding and return value usage with many-to-many callsite-callee type-based matching to enforce CFI. 
Unfortunately, TypeArmor often vastly under/over-estimates argument count, resulting in false negatives or positives. 
Similarly, TypeSqueezer~\cite{Lin2023TypeSqueezer} relies on runtime analysis to refine argument counts and enrich function signatures by distinguishing between data and pointers. 
However, TypeSqueezer's reliance on dynamic execution necessitates a comprehensive test suite for effective performance.
BPA~\cite{BPA} introduces a memory block model to enhance points-to analysis by eliminating offset tracking, thereby improving the scalability of VSA. 
Nevertheless, BPA's memory block model makes strong assumptions, such as relying on heuristics for boundary recovery based \emph{solely} on calling conventions in the x86-32 architecture, 
which impacts their performance as evident in their lower F1 score when compared to ML approaches in Table~\ref{table:CurrentApproaches}.
All three tools, unfortunately, face challenges in achieving a favorable balance of precision, recall, and scalability. For instance, while value-set analysis (VSA) as used in BPA is a powerful technique that intrinsically tracks data and code references (xRefs) to resolve potential targets, it can suffer from scalability issues and over-approximation due to the state explosion problem in large, complex binaries. Our ML-based approach is complementary; by representing xRefs and control flow in a graph structure, \mytool learns contextual patterns that may be difficult to capture in traditional abstract domains.

\subsection{AI-based Indirect Call Resolution}

While AI/ML techniques have been actively applied to address several challenges in binary analysis, such as binary diffing~\cite{DeepBinDiff, pei2020trex,he2024code}, debug symbol recovery~\cite{debin}, type prediction~\cite{pei2021stateformer}, and function name prediction~\cite{david2020neural,jin2022symlm}, current solutions for AI-based binary analysis are not directly applicable to icall resolution. 
These solutions predominantly focus on the semantics of assembly instructions, neglecting the crucial data and address xRef information present in binaries. 
We compare \mytool with peer works in Table~\ref{table:CurrentApproaches}.
Moreover, current ML approaches utilize standard embedding schemes, which can lead to reduced granularity and information loss (refer to $\S$\ref{sec:challenges}), potentially impacting a model's ability to make informed decisions. Despite these shortcomings, ML binary analysis tools continue to outperform traditional approaches such as BPA~\cite{BPA} and TypeArmor~\cite{TypeArmor}. 
We believe that improvements in icall prediction accuracy require the integration of xRef information and better binary code representation. 
Next, we discuss two recent papers closely related to our work.


\vspace*{2pt}
\noindent \textbf{Callee~\cite{Callee} }
Callee infers icall targets by leveraging Siamese neural networks~\cite{bromley1993signature} and applying transfer learning on a limited indirect call dataset. 
While Callee reports an F1 score of 94\%, our initial evaluation with their open-source code and pre-trained model yielded a much lower F1 score of 43.9\%. 
After re-implementing Callee's training procedure and fine-tuning it with our collected icallsite-callee pairs, we improved Callee's F1 score to 89.9\%.
A key limitation of Callee is its dependence on hardware tracing and emulation for ground truth, which introduces false negatives and provides an incomplete and small dataset of indirect calls.  
In contrast, our approach utilizes compiler-level type analysis to capture indirect calls without the need of complex testing suites, allowing us to obtain a significantly larger and more diverse dataset.
Additionally, Callee fails to utilize data section and xRef information present in binaries, which are crucial for enhancing semantic information retention. 
Due to this lack of rich information, Callee resorts to a ``loose'' tokenization approach to handle out-of-vocabulary (OOV) issues, leading to information loss and poor binary code representation (further described in $\S$\ref{OOV}). 
By integrating additional useful information such as xRefs, our model enhances learning and inference capabilities, offering a more comprehensive solution for icall target prediction.

\vspace*{2pt}
\noindent \textbf{AttnCall~\cite{AttnCall2024} }
AttnCall utilizes the attention mechanism~\cite{vaswani2017attention} to establish the contextual relationship between callsites and callees.
However, their approach has several significant limitations
Firstly, AttnCall's authors have not released a pre-trained model or a corresponding dataset, hindering reproducibility efforts. 
Second, the model's reliance on random slicing reduces accuracy as it fails to capture the multifaceted nature of a function's behavior when constrained to a linear representation.
Moreover, AttnCall has not been tested on icalls; it is exclusively trained and tested on direct calls (dcalls). Consequently, their reported F1 score pertains to a fabricated dcall dataset rather than actual icall targets.
They use fabricated functions as negative samples for training and testing, trivializing the model's task of distinguishing true targets from fabricated ones, artificially inflating its F1 score.
Like Callee, AttnCall uses a conventional tokenization approach for OOV mitigation in their NLP model. 
This approach, however, results in the loss of critical xRef information, thereby preventing accurate training. Our model addresses this issue by representing binaries as graphs and 
incorporating xRefs as new edges (see $\S$\ref{sec:BinaryRep} for details on improved binary code representation).
To highlight AttnCall's shortcomings, we simulate their training methodology to demonstrate that icall prediction exclusively from dcalls is unrealistic.

\subsection{Compiler-level CFI} 


There is no established ``ground truth''in the realm of indirect control flow resolution tasks. 
Dynamic methods, which gathers execution traces via binary instrumentation or hardware tracing, cannot ensure complete path coverage, resulting in false negatives.  
Conversely, static methods, e.g. value set analysis, struggle to avoid false positives.
A primary challenge in binary code analysis is the loss of crucial information, such as types and prototypes, during compilation. 
This loss significantly complicates binary-level analysis and diminishes its accuracy compared to compiler-level analysis. 
Compiler-level analysis, with full access to source code, offers the most comprehensive insights into a program's semantics.
Therefore, \textit{our objective is to develop a binary-level icall target predictor that can replicate the capabilities achieved by compiler-level analysis}. 

The current state-of-practice tool for icall target resolution at the compiler level is LLVM-CFI~\cite{clang-cfi,llvm-api}, which is widely applied in real-world applications such as the Linux Kernel. 
Notably, the authors of Callee acknowledge that LLVM-CFI continues to outperform their model, motivating our decision to leverage an LLVM-CFI-based analysis tool to improve upon existing work. 
However, LLVM-CFI’s soundness remains a point for improvement.
Two notable works address the limitations of Clang-CFI. 
The first, TypeDive~\cite{lu2019does}, enhances indirect call target refinement accuracy through multi-layer type analysis and the incorporation of additional type hierarchy structure information. 
The second, TyPro~\cite{bauer2022typro}, suggests that collecting type propagation during compile time to capture indirect call targets can reduce the false negative rate more effectively than typical LLVM-CFI, rather than directly matching type information.
As minimizing false negatives is our priority, we select TyPro as 
our icallsite-callee collection mechanism. 
However, TyPro does not address the challenge of mapping source information to the binary level, as it was beyond the intended scope. 
Details on how we modify TyPro and the
linker to overcome this challenge are described in $\S$\ref{sec:TyProMod}.

\subsection{Graph Neural Networks}


Given that programs can be represented as graphs, GNNs are a natural choice for our task~\cite{david2020neural,CodeArt}.
This decision allows us to preserve more useful information, such as xRefs, that standard data structures may overlook. 
Since homogeneous graphs cannot distinguish between different types of edges, such as control flow edges and xRef edges, 
we utilize a heterogeneous graph to accurately represent their relationships within binary code. 
Moreover, traditional message-passing GNNs often perform poorly when positional information of nodes is missing, particularly in tasks like cycle detection~\cite{murphy2019relational,dwivedi2020benchmarking,dwivedi2022graphPE}. 
To address this issue, node positional encoding has been proposed, demonstrating substantial performance improvements across various tasks, such as social network analysis and molecule analysis~\cite{dwivedi2020benchmarking}.

\vspace*{2pt}
\noindent \textbf{Heterogeneous Graph }
A heterogeneous graph comprises nodes and edges of multiple types, representing diverse entities and relationships rather than a uniform set. 
This structure allows for a more comprehensive representation of information, such as integrating code and data or control flow and cross-reference relationships within a single graph. 
Recent research in heterogeneous graph analysis~\cite{RGCN} has focused on developing new techniques for graph embedding, which aims to map the nodes in the graph to a low-dimensional space while preserving the graph structure and the heterogeneity of nodes and edges~\cite{hamilton2017inductive}.


\vspace*{2pt}
\noindent \textbf{Graph Positional Encoding }
Graph positional encoding (GPE) is a technique used in graph neural networks to incorporate the position information of nodes in a graph. The basic idea behind GPE is to assign a unique position vector to each node in the graph, which captures its relative position with respect to other nodes in the graph~\cite{dwivedi2020benchmarking,dwivedi2020generalization}. By incorporating position information into the node representations, GPE can help the model better understand the structure of the graph and capture the local and global relationships between nodes.



\subsection{Assembly Instruction Embedding}
A binary program can be represented either as machine code or assembly language, both of which are challenging to use in machine learning applications. To address this, word embedding techniques from the NLP domain become ideal. 
An embedding translates high-dimensional text into low-dimensional vectors, grouping instructions with similar properties. BERT~\cite{devlin2018bert} is the state-of-the-art model for word embeddings. Additionally, several works have focused on embedding binary programs, such as Instruction2Vec~\cite{lee2019instruction2vec}, InnerEye~\cite{zuo2018neural}, Asm2Vec~\cite{ding2019asm2vec}, and PalmTree~\cite{li2021palmtree}. Among these, PalmTree is a general-purpose instruction embedding model based on BERT that outperforms other similar models. It is a self-supervised model capable of capturing the relationships between instructions. In our work, we utilize PalmTree for instruction embedding processing.

\begin{figure*} [t]
	\centering
	\includegraphics[width=0.84\textwidth,angle=0,scale=1.0]{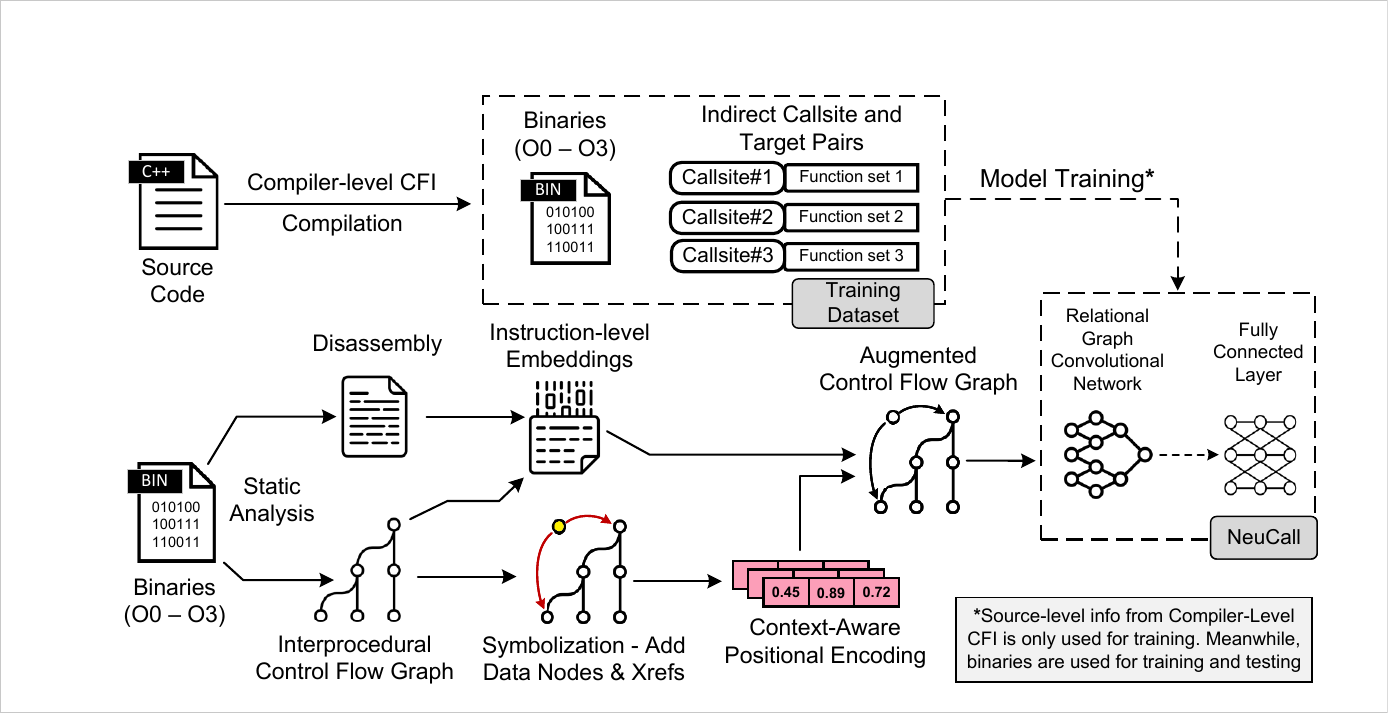}
	\vspace{-1mm}
	\caption{Data preprocessing overview: prepares input for relational graph convolutional network.}
	\label{fig:overview2}
	\vspace{-5mm}
\end{figure*}

\section{Motivation and Overview}\label{sec:overview2}
This section outlines the key challenges in prior AI-based binary analysis work,  particularly concerning information loss and binary code representation, which we aim to address with \mytool.


\subsection{Challenges}\label{sec:challenges}
Previous efforts in AI-based binary analysis face several pitfalls related to information loss, including missing data and code xRef information, OOV issues, and limited path coverage. 
Figure~\ref{fig:infoloss} illustrates the extent of information loss, such that data section information concerning icalls and address xRef are stripped away during tokenization.

\vspace*{2pt}
\noindent \textbf{Data vs. Code }
Many previous works~\cite{pei2020trex,pei2021stateformer,jin2022symlm} only use
disassembled instructions from the ``.text'' section to represent a binary file in machine learning. 
However, the data section of a binary file contains additional vital information such as function pointers, jump table targets, initial variable values, and string information. 
Cross-reference information such as function pointers and jump table targets are crucial for constructing control flow and have been neglected in previous works.
We aim to use underutilized xRef information to improve the prediction accuracy of our model.

\vspace*{2pt}
\noindent \textbf{Out of Vocabulary (OOV) }
\label{OOV}
The vocabulary of a language model typically consists of a fixed set of words defined during the training process, limited to the most frequent words in the training data. OOV words are a common problem in NLP tasks, 
referring to words not included in a language model's vocabulary. OOV words can cause errors or inaccuracies in NLP tasks. In binary analysis scenarios, addresses and symbols are common examples of OOV challenges.
A common mitigation strategy is to replace uncommon numbers or symbols with special tokens, such as ``[num]'' or ``[sym]''~\cite{lee2019instruction2vec, li2021palmtree}. This approach, known as strict tokenization, assigns a specific token to each type of number or symbol. Unlike previous approaches, Callee~\cite{Callee} attempts to encode an unlimited number of symbols into a fixed set of tokens. This method, referred to as loose tokenization, aims to preserve more data-flow information compared to strict tokenization. To address the OOV challenge, Callee sets a hyperparameter \texttt{N} to 10, creating a finite corpus of symbols (e.g., ``[addr1, ..., addr10]'').
In binary analysis, replacing all numeric addresses with a single token (e.g., [addr]) eliminates semantic distinctions between distinct addresses, leading to information loss. While Byte Pair Encoding (BPE)~\cite{bostrom-durrett-2020-byte} might theoretically mitigate the OOV issue, our experiments (see Appendix~\ref{sec:BPE}) show that BPE‐based subword embeddings still fail to encode address–address relationships adequately.


\vspace*{2pt}
\noindent \textbf{Representation of Binary Code }
Natural language processing (NLP) models typically assume a linear relationship among the inputs, which results in the loss of critical control flow information. Existing approaches, as described in~\cite{Callee,pei2020trex,pei2021stateformer,jin2022symlm}, 
typically rely on linear inputs that either represent the function's semantic regarding paths or just the address order.
However, a path or linear input is not suitable for accurately representing binary code. 
Consequently, tools that collate several paths or execution traces fail to provide an accurate representation. 
Moreover, using a linear representation for binary code leads to the loss of crucial xRef information.

\begin{figure*} [t]
	\centering
	\includegraphics[width=0.9\textwidth,angle=0, scale=1.0]{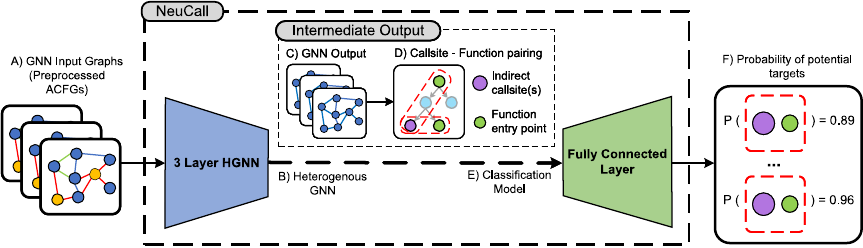}
	\caption{Overview of the \mytool architecture for training and testing, where the \textbf{input} is A) the constructed Augmented CFGs (ACFGs) after preprocessing, and the \textbf{output} is F) the probability of a function being the target of an indirect callsite.} 
	
	\label{fig:GNN-Pipe}
	\vspace{-5mm}
\end{figure*}

\subsection{Key Insight: Improved Binary Representation}
\label{sec:BinaryRep}
As discussed in $\S$\ref{sec:challenges}, encoding and learning arbitrary numbers using NLP embeddings is a challenging task for deep learning models. 
In the context of binary analysis, distinct numerical values, such as instruction addresses, have the unique property of being references. 
These references can naturally be represented as edges within a graph structure.
To improve the representation of binary programs, we augment traditional CFGs by incorporating reference edges and introduce data nodes to link the CFG with these edges. 
This enhanced representation, capturing both control flow and data references, provides deeper insights on semantic structure and enables more accurate inferences on indirect control flow.

\noindent \textbf{Encoding Augmented CFGs with GNNs }
A control flow graph provides a static representation of a program's workflow. He and Lin et al.~\cite{he2024code} highlight the issue that binary code cannot be treated similarly to natural language tasks due to significant structural differences between assembly code and natural languages. While their observations align with ours, they do not consider address analysis and additional semantic information, which are crucial for icall prediction. 
Nero~\cite{david2020neural} is another work that shares similar design choices, aiming to provide CFGs with contextually rich information to predict function names more accurately. 
However, Nero focuses on constructing augmented call graphs via pointer-aware slicing to enhance call sites with abstracted/concrete values,
rather than directly utilizing xRef information.





When encoding CFGs for NLP-related tasks, crucial pointer and address information is often lost. Therefore, it is essential to utilize the appropriate model, GNNs, to better represent CFGs. 
Furthermore, incorporating extra semantic information into CFGs can significantly enhance the model's ability to infer icall targets.
To achieve this, we integrate the binary's data section and code/data xRefs into CFGs after performing symbolization~\cite{angr2016artofwar, pang2021sok}. 
This approach, often overlooked in existing works, also addresses the binary code's OOV issues, potentially improving the accuracy of icall resolution.

\subsection{\mytool Workflow}

The workflow of \mytool consists of four major steps.

\vspace*{2pt}
\noindent \textbf{I. Training Data Collection }  Given source code, we utilize compiler-level CFI (see Figure~\ref{fig:overview2}) to collect indirect callsites and targets to enhance \mytool's learning ability. These icallsite-callee pairs serve as \mytool's training input.

\vspace*{2pt}
\noindent \textbf{II. Graph Structure Construction } In order to construct the augmented CFG inputs for \mytool, as shown in Figure~\ref{fig:overview2}, we employ symbolization to capture code/data xRefs to enrich CFGs by adding new xRef edges and data nodes. 

\vspace*{2pt}
\noindent \textbf{III. GNN Node Embedding } 
Simultaneously, we apply Laplacian position encoding to augmented CFGs while encoding assembly instructions using PalmTree~\cite{li2021palmtree} to construct the final GNN node embeddings,
which further serve as \mytool's input for training and testing.

\vspace*{2pt}
\noindent \textbf{IV. Heterogeneous Graph Neural Networks } 
\mytool's heterogeneous GNNs are trained using augmented CFGs and icallsite-callee pairs. During the testing stage, given the embeddings of the augmented CFGs as input, \mytool outputs the probability of a function being the target of an icall. 
The model architecture is summarized in Figure~\ref{fig:GNN-Pipe}.

\section{System Design}\label{sec:model}

This section presents the detailed design of \mytool.
To enhance the quality of our training data, we modify an LLVM plugin, TyPro~\cite{bauer2022typro}, to collect and map indirect callsites and targets. We then utilize symbolization to preserve cross-reference information to further enhance the quality of CFGs, which we call augmented CFGs (ACFG). We additionally apply existing GNN optimizations such as Position Encoding. \mytool, which consists of a GNN and a fully connected layer, is trained on these preprocessed ACFGs and the collected icallsite-callee pairs.





\begin{figure*} [t]
	\centering
	\includegraphics[width=0.95\textwidth,angle=0, scale=1]{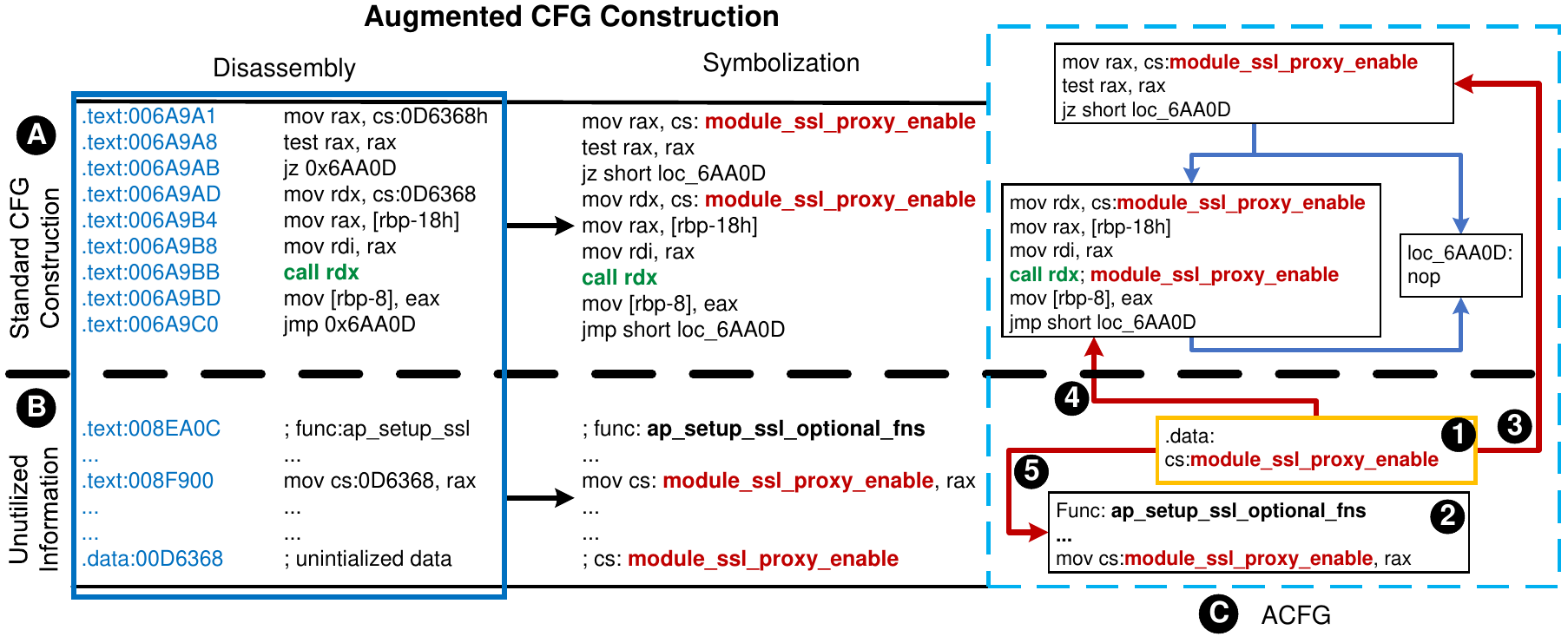}
	\vspace{-2mm}
	\caption{Augmented control flow graph construction via symbolization and cross-references. This disassembly code snippet is picked from Apache HTTP Server's ap\_ssl\_bind\_outgoing function.}
	\label{fig:example}
	\vspace{-5mm}
\end{figure*}

\subsection{Toolchain Customization} 
\label{sec:ground_truth_collection}

\label{sec:TyProMod}


Before utilizing TyPro, modifications are necessary to accurately map source-level information to the binary level, as failing to do so could lead to inaccuracies in constructing our training data. 
Initially, obtaining address information poses a challenge because addresses are assigned at the linker stage, and TyPro, being a compiler-level tool, lacks the capability to provide direct address locations for icalls. To solve this, we modified the toolchain:
\begin{itemize}
\item During compilation, our modified TyPro plugin inserts unique \textbf{labels} to mark the locations of indirect callsites and targets.
\item During linking, we use a custom linker script that resolves these labels into their final \textbf{addresses}.
\end{itemize}
This process creates a map from source-level information to the final binary addresses. Compared to the strategies used in peer works~\cite{Callee,AttnCall2024}, our customized toolchain allows us to achieve a more accurate and complete representation of icallsites and their corresponding targets, 
thereby providing a reliable training dataset for future analysis.

\subsection{Graph Structure Construction}
\label{sec:preprocessing}
\noindent \textbf{Symbolization}
To assist in discovering data-related edges, such as data-to-code reference edges, we leverage symbolization techniques. 
Symbolization is the process of identifying references among immediate values in disassembled code~\cite{wang2015reassembleable}, 
and it is essential for preserving both xRefs and data information without introducing OOV issues. 
Wang et al.~\cite{wang2015reassembleable} suggest that if all symbols can be resolved, the disassembled assembly can be recompiled into a functional binary.
Thus, with address information, fully symbolized code and data can comprehensively represent the binary. 
However, this technique is not without flaws, as it relies on heuristics and assumptions that potentially introduces false positives and false negatives. 

Nevertheless, symbolization remains relatively accurate for many practical applications in binary analysis, as its methods are designed to generalize well across common reference patterns and address formats. In fact, Pang et al. demonstrates that symbolization techniques exhibit high precision~\cite{pang2021sok}.
Furthermore, symbolization has been proven that it can guarantee soundness in a recent application, specifically for lifting x86 binaries~\cite{verbeek2024verifiably}. 
Angr has demonstrated high precision and recall rates (99.9\% and 99.7\%, respectively)~\cite{pang2021sok}, making it a reliable choice. Therefore, we utilize angr~\cite{angr2016artofwar} to perform symbolization, capturing the necessary xRef and data information to enhance CFGs.

\vspace*{2pt}
\noindent \textbf{Augmenting xRef Information into CFGs }
To capture more details into a program's behavior with respect to icalls, constructing fine-grained CFGs is essential. 
While various parts of a binary file can be considered, we hypothesize that incorporating five specific types of information will suffice to achieve our goal. 
Thus, we augment the CFG with the following node and edge feature types: \textit{data node}, \textit{code-to-data (c2d) reference edge}, \textit{data-to-data (d2d) reference edge}, \textit{data-to-code (d2c) reference edge}, and \textit{code-to-code (c2c) reference edge}.


A data node is created and added to the CFG after symbolizing all to-data xRefs. We divide the data sections with these to-data labels into individual data nodes. 
For c2d references, we add a data-being-referenced edge from the referenced data node to the corresponding referencing basic block node. 
For d2d references, similar to c2d references, we add a data-being-referenced edge from the referenced data node to the referencing data node. 
For c2c references, which differ from conventional control flow transfer edges, a distinct edge is created between the reference address within the code block and the targeted code block, explicitly denoting reference relationships outside the usual control flow.
Finally, for d2c references, we add a code-being-referenced edge from the referenced basic block node back to the referencing data node, capturing instances where data elements reference executable code. This enriched CFG structure provides a more comprehensive view of a program's semantics.

In Figure~\ref{fig:example}, we present a running example of constructing an augmented control flow graph (ACFG) using previously overlooked information. 
The standard CFG construction method, represented by \circled{A} in Figure~\ref{fig:example}, omits xRef information. 
Conversely, if xRefs are integrated into the CFG, as shown in \circled{B} and \circled{C}, the hidden memory read/write relationships with icall targets (e.g., ``call rdx'' in Figure~\ref{fig:example}) become evident.

The provided disassembly code snippet, obtained from the ``ap\_\allowbreak ssl\_\allowbreak bind\_\allowbreak outgoing'' function in the Apache HTTP Server, shows that the target of ``call rdx'' is loaded from the data section at ``cs:0D6368h.'' 
Referring back to Figure~\ref{fig:infoloss}, the conventional approach typically excludes the data section and replaces addresses with special tokens like ``[addr]''~\cite{pei2021stateformer,pei2020trex}. 
As a result, encoding this function without incorporating data and addresses would cut off the connection between the icallsite and its callee.

Due to the limitations of standard CFGs, which lack xRef information and encoding schemes that distinguish unique addresses, we aim to construct an ACFG to retain more semantic information. 
In the final ACFG, as illustrated in \circled{C}, we demonstrate how c2d reference edges (\circled{3} and \circled{4}) originating from the standard CFG reveal a connection to a new data node \circled{1}. 
This data node serves to connect the code node \circled{2} via c2d reference edge \circled{5}, enabling us to identify the setup function required for preparing the target of ``call rdx.''

\subsection{GNN Node Embedding}
\noindent \textbf{Position Encoding in GNNs }
Node positional encoding annotates the structural position of nodes within a graph. In our work, we employ Laplacian eigenvectors as node positional features, utilizing Laplacian Positional Encoding (Laplacian PE)~\cite{belkin2003laplacian}. 
Although Laplacian PE is not specifically designed for heterogeneous graphs because it does not account for the relationships between different types of nodes and edges, 
it possesses the notable strength of being network-agnostic. This means it encodes positional information directly into the node features.
In our approach, we first transform the heterogeneous graph (our ACFG) into a homogeneous graph. We then calculate the positional encoding and incorporate it back into the heterogeneous graph. 
This process allows us to leverage the advantages of Laplacian PE while accommodating the complexities of heterogeneous graph structures.




\vspace*{2pt}
\noindent \textbf{Code Node Embedding}
To embed assembly code into our model, we adopt the methodology used by PalmTree~\cite{li2021palmtree}, a BERT-based assembly language model. 
PalmTree treats each instruction as a sentence, tokenizes it, and employs BERT's Masked Language Model to predict missing tokens within the instruction. 
It infers instruction semantics by predicting the co-occurrence of instructions within a sliding context window in the control flow.
As a result, this instruction embedding scheme can capture more semantic information. 

To provide the GNN with a global sense of location, we supplement the instruction embeddings with the normalized address of the basic block itself. This feature is distinct from the structural Laplacian PE and serves to help the model differentiate between code segments that may be structurally similar but reside in different parts of the binary, such as library code versus main application code.
Specifically, for each code node (basic block), we add its {normalized} address and corresponding {normalized} function address, and append this information to the initial instruction embedding to form our ``code node embedding.''



\vspace*{2pt}
\noindent \textbf{Data Node Construction} We preserve the address information of the data block and extract the reference as a graph edge, as illustrated in Figure~\ref{fig:example}. 
Similar to code node embedding, we use a data block's normalized address as our ``data node embedding.''
                  
\subsection{Heterogeneous Graph Neural Networks}
The primary function of a GNN~\cite{scarselli2008graph} is to propagate information across the graph structure, enabling the model to extract useful features and make predictions. 
Our model training and testing pipeline is illustrated in Figure~\ref{fig:GNN-Pipe}. 





As shown in Figure~\ref{fig:GNN-Pipe}, the GNN input graphs of \mytool is \textbf{A)} our preprocessed ACFGs (see Figure~\ref{fig:overview2}), which transforms into a heterogeneous graph with various node and edge types. 
To preserve the relational connections between code and data, we utilize \textbf{B)} a heterogeneous graph neural network, specifically {a relational graph convolutional network}.
Our ACFG is defined as \[G = (V,E,R,T)\] 
$v_c$ is the basic block nodes, $v_d$ is the data nodes, where \[V=\{v_c,v_d\}\] 
We separate the call edges, $r_c$, with the other control flow transfer edges, $r_t$. The $r_{cc}$ denotes c2c reference edges. The $r_{cd}$ denotes c2d reference edges. The $r_{dc}$ denotes d2c reference edges. The $r_{dd}$ denotes d2d reference edges. Thus, we define $R$ as the set of all reference edges, where \[R=\{r_c,r_t,r_{cc},r_{cd},r_{dc},r_{dd}\}\]
Edges are defined with the source node, relation edge type, and destination nodes. \[(v_i, r, v_j)\in E\]

In the message passing phase, information is propagated across the graph structure through a series of local operations. Each node aggregates information from its neighbors, which is then combined with the node’s own features to produce a new feature vector. This process repeats iteratively until information has propagated across the entire graph.
The message passing function is defined as \[h^{(l+1)}_v=f(\sum_{r\in R}\sum_{u\in N^r_v}\frac{1}{c_{v,r}}W^{(l)}_rh^{(l)}_u+W^{(l)}_0h^{(l)}_v)\]
The feature representation at layer $l$ is $h^l_u$ where $W^l_r$ is the weights at layer $l$. $c_{v,r}$ is normalized by node degree of the relation.


\vspace*{2pt}
\noindent\textbf{Classification Model } As shown in Figure~\ref{fig:GNN-Pipe}, we utilize a fully connected layer \textbf{E)} as our classification model. The input for this layer is \textbf{D)} callsite-function embedding pairs derived from \textbf{C)} the GNN output,
specifically the final node representations from the GNN's readout phase, denoted as {$h$} (the last layer's output). Each embedding pair is used to predict whether an icall, $r_c$, exists between the icallsite $v_i$ and a potential callee function's entry point block. The output of the model \textbf{F)}, $f_{r_c}$, represents the probability 
that the callee's entry block is indeed the target of the icall site.



\vspace*{2pt}
\noindent\textbf{Loss Function}
$h_t$ denotes a true target callee. $h_f$ denotes a false target.
We aim to minimize the loss function, defined as: \[l=-log f_{r_c}(h_i,h_t) - log(1- f_{r_c}(h_i,h_f))\]

We train both GNN and the classification model with our labeled icallsite-callee pairs simultaneously.

\section{Evaluation} 

As Callee~\cite{Callee} has already shown, a refined indirect call target set can significantly enhance downstream security tasks, such as binary similarity detection and hybrid fuzzing. Therefore, the primary objective of our evaluation is to demonstrate that our model provides substantial improvements in indirect call target refinement. We evaluate \mytool from four perspectives: 
1) a hyperparameter tuning experiment to determine the optimal values for our model's hyperparameters;
2) an ablation study to assess the performance contribution of each model component; 
3) a comparative study against the current state-of-the-art solutions; 
4) performance across different optimization levels;
and 5) a case study of security hardening application.


\begin{table}
	\caption{Training dataset statistics. The values in Column~4$\sim$8 represent the corresponding quantities.}
	\label{table:datasets}
    \vspace{-1mm}
	\centering
	\fontsize{9pt}{9pt}\selectfont
    \resizebox{0.49\textwidth}{!}{
	\begin{threeparttable}
		\begin{tabular}{lrcrrrrr}
			
			\topline
			\headcol 

			&
			\thead{\textbf{\# of Total} \\ \textbf{Binaries}}
            &
			\thead{\textbf{Access to} \\ \textbf{Binaries?}}
			&
			\thead{\textbf{Usable} \\  \textbf{Binaries\tnote{a}}}
			&
			\thead{\textbf{Funcs}}
			&
			\thead{\textbf{Ind. } \\ \textbf{calls}}
			&
			\thead{\textbf{Ind.} \\ \textbf{call pairs}\tnote{c}}
            &
			\thead{\textbf{xRefs}}

			\\ \hline
			\raisebox{-0.5ex}[-0.5ex]{GitHub} & \raisebox{-0.5ex}[-0.5ex]{$3,154$} & \raisebox{-0.5ex}[-0.5ex]{\greencheck}& \raisebox{-0.5ex}[-0.5ex]{596} & \raisebox{-0.5ex}[-0.5ex]{98K} & \raisebox{-0.5ex}[-0.5ex]{6K} & \raisebox{-0.5ex}[-0.5ex]{24K} & \raisebox{-0.5ex}[-0.5ex]{169K} 
            \\
			\rowcol
			\raisebox{-0.5ex}[-0.5ex]{Arch} & \raisebox{-0.5ex}[-0.5ex]{$22,013$} & \raisebox{-0.5ex}[-0.5ex]{\greencheck} & \raisebox{-0.5ex}[-0.5ex]{2084} & \raisebox{-0.5ex}[-0.5ex]{252K} & \raisebox{-0.5ex}[-0.5ex]{30K} & \raisebox{-0.5ex}[-0.5ex]{680K} & \raisebox{-0.5ex}[-0.5ex]{699K} 
            \\
					\bottomlinec
					\raisebox{-0.5ex}[-0.5ex]{Total} & \raisebox{-0.5ex}[-0.5ex]{$25,167$} & \raisebox{-0.5ex}[-0.5ex]{\greencheck} & \raisebox{-0.5ex}[-0.5ex]{2680} & \raisebox{-0.5ex}[-0.5ex]{350K} & \raisebox{-0.5ex}[-0.5ex]{36K} & \raisebox{-0.5ex}[-0.5ex]{704K} & \raisebox{-0.5ex}[-0.5ex]{868K} 
                    \\ 
        \bottomlinec
        \rowcol
        \raisebox{-0.5ex}[-0.5ex]{Callee\tnote{b}} & \raisebox{-0.5ex}[-0.5ex]{268} & \raisebox{-0.5ex}[-0.5ex]{\rxmark} & \raisebox{-0.5ex}[-0.5ex]{0} & \raisebox{-0.5ex}[-0.5ex]{28K} & \raisebox{-0.5ex}[-0.5ex]{31K} & \raisebox{-0.5ex}[-0.5ex]{50K} & \raisebox{-0.5ex}[-0.5ex]{N/A} 
        \\
			%
			\bottomlinec
			
		\end{tabular}
		\begin{tablenotes}
			\small
			\item[a] Binaries that either contain no indirect calls or cannot be analyzed by TyPro are excluded from our evaluation dataset. For \mytool, the average number of xRefs per function is 557.5.
            \item[b] Callee authors do not provide the raw binaries, prohibiting accurate replication of their model's performance.
            \item[c] The valid icallsite-target pairs in the training dataset without invalid pairs. 
		\end{tablenotes}
		
	\end{threeparttable}
 }
\vspace{-5mm}
\end{table}


\subsection{Setup}
We performed all experiments on a Windows 10 machine with subsystem for Ubuntu 20.04 LTS. 
The machine has an Intel(R) Xeon(R) Gold 6242R CPU @ 3.10GHz, two NVIDIA RTX A6000 GPUs, and 512 GB of RAM. Python 3.9.1, DGL 0.9.1, and PyTorch 2.2.0 are used to build \mytool.

\vspace*{2pt}
\noindent \textbf{Dataset }\label{data_section}
We compile programs with TyPro~\cite{bauer2022typro}, an LLVM-CFI based analysis tool, from two different sources to construct our dataset of icalls and icall targets necessary for training and evaluating our model. 
The first source is from public GitHub repositories, which we employ GitHub Commit Crawler to collect programs. 
The second source is from the Arch User Repository which contains a popular and diverse set of large and small programs. Furthermore, for our AUR source, we build binaries for optimization levels O0-O3.
Table~\ref{table:datasets} shows the number of binaries, functions, icalls, and other statistics obtained from our training data collection process.
In total, 596 usable binaries were compiled by LLVM 10 from GitHub and 2,084 from the Arch User Repository~\cite{Arch}.  Overall, it takes approximately 67 hours to compile 2,084 usable binaries from AUR repository and approximately 7 days to compile 596 usable binaries from GitHub.
Except for the cross-optimization level evaluation, we adhered to the default build configurations provided by each project and did not enforce static linking or whole-program optimization. Consequently, our dataset comprises a realistic mix of statically and dynamically linked executables, reflecting typical software distribution. \mytool operates on these individual binary files as-is, a scenario representative of real-world binary analysis where external libraries may not be available for simultaneous analysis.

To avoid confusion, it is important to clarify that Callee's 50K ``pairs'' do not represent verified one-to-one icallsite–callee matchings. Rather, they are derived from the Cartesian product of all icall sites and address-taken functions, leading to numerous spurious candidates.
In contrast, we report 36K indirect calls and a total of 704K icallsite–callee pairs, all validated using TyPro's type propagation. To the best of our knowledge, this constitutes the largest verified icallsite–callee dataset to date. This scale is made possible by adopting a more practical and cost-effective strategy that collects icallsite–callee pairs directly during compilation.

\vspace*{2pt}
\noindent \textbf{Limitations in Callee Dataset Reproduction }
The authors of Callee provide a pre-processed dataset used for training their model; however, we are unable to leverage this dataset as the original binaries are not included, where data information is already lost in their pre-processed dataset. 

\vspace*{2pt}
\noindent \textbf{Sampling Process }
For each program, we employ a balanced sampling strategy similar to Callee. This involves randomly selecting an equal number of functions from the binary that are not among the callsite's true targets. For example, if a program contains 100 functions and an indirect call has four positive targets (Func01, Func03, Func22, Func44), then four additional functions (e.g., Func02, Func15, Func28, Func35) are randomly selected as negative examples. In other words, any function not in the true target set is eligible as a negative sample. This approach ensures the training and evaluation datasets contain a balanced mix of positive and negative examples, while introducing variability across epochs to mitigate overfitting~\cite{balancedData_1,balancedData_2,balancedData_3}.

By employing this methodology, the training and evaluation datasets were optimized for machine learning purposes, thus improving the overall performance and accuracy of the models.
We randomly split the data set into three sets. 80\% for training, 10\% for validation, and 10\% for testing. 

\vspace*{2pt}
\noindent \textbf{Project-Level Dataset Splitting }
To prevent information leakage and ensure our model generalizes to unseen code, we split our dataset at the project level. All binary files compiled from the same source project were assigned to a single data split (i.e., either all in training, all in validation, or all in testing). This strict separation ensures that the model is evaluated on its ability to analyze binaries from projects it has not been trained on, offering a more rigorous and realistic measure of \mytool's generalization capability compared to prior work that may perform splits at the pair or binary level.~\cite{Callee,AttnCall2024}


\vspace*{2pt}
\noindent \textbf{Models and Hyperparameters }
The GNN model utilized in this study is a three-layer RGCN~\cite{RGCN}, with a three linear fully connected layer link predictor. The model was trained using DGL front-end~\cite{DGL} on top of PyTorch 2.2.0. During training, the input to each batch was a program's augmented CFG, which was learned and tested across all icallsite-callee pairs of that program. Our instruction embedding size is set to 70 instructions. 
The network was trained with a learning rate of 0.001, and a hidden layer feature size of 512 and RGCN layer depth of 3. A dropout rate of 0.2 was also utilized to prevent overfitting. 


\noindent \textbf{Evaluation Metrics }
To evaluate the model's overall performance, we utilized commonly used metrics, including Precision, Recall, and F1-Score. These metrics were calculated based on the number of True Positives (TP), True Negatives (TN), False Positives (FP), and False Negatives (FN) generated by the model's predictions. For our model, a false positive would be a function that is not a possible target of an icall. A false negative would be a true icall target being mislabeled as false. 
In addition to Precision, Recall, and F1-Score, we also report the Area Under the Receiver Operating Characteristic Curve (\textbf{AUROC}) metric as a measure of the model's performance. AUROC is a commonly used metric for binary classification models which measures the model's ability to distinguish between positive and negative examples across a range of classification thresholds.

\subsection{Hyperparameter Tuning}
\label{sec:Hyper_Eval}

We conducted two experiments to identify the parameter configurations that yield the most optimal prediction results. The first experiment aims to determine the optimal GNN layer depth and hidden feature size. The second experiment assesses various embedding lengths to evaluate their impact on the model’s performance. Each hyperparameter experiment is based on Setting 10, as defined in the last row of Table~\ref{table:ablation}), where the following GNN features are enabled: reverse edges, data nodes, data reference edges, code reference edges, function nodes, call edges, and position encoding.



\begin{table} [t]
	\caption{Layer size impact on model. Instruction embedding length and hidden feature size of 70 and 512 respectively.}
    \vspace{-1mm}
	\centering
    \resizebox{0.8\columnwidth}{!}{
	\begin{tabular}{ccccc}
		\topline
		\headcol 
		
		\thead{\# RGCN \\ Layers}  
		
		&
		F1
		&
		Precision
		&
		Recall
		&
		AUROC
		
		\\ \hline
		\raisebox{-0.5ex}[-0.5ex]{1} &  \raisebox{-0.5ex}[-0.5ex]{92.2\%} &  \raisebox{-0.5ex}[-0.5ex]{92.5\%} &  \raisebox{-0.5ex}[-0.5ex]{91.9\%} & \raisebox{-0.5ex}[-0.5ex]{96.3\%} \\ 
		\rowcol
		\raisebox{-0.5ex}[-0.5ex]{2}	&		\raisebox{-0.5ex}[-0.5ex]{95.1\%}	&		\raisebox{-0.5ex}[-0.5ex]{94.0\%}	&	\raisebox{-0.5ex}[-0.5ex]{96.2\%}	&	\raisebox{-0.5ex}[-0.5ex]{98.0\%} \\

		\raisebox{-0.5ex}[-0.5ex]{\textbf{3}}	&		\raisebox{-0.5ex}[-0.5ex]{\textbf{95.2\%}}	&			\raisebox{-0.5ex}[-0.5ex]{\textbf{97.1\%}}	&	\raisebox{-0.5ex}[-0.5ex]{\textbf{93.3\%}}&	\raisebox{-0.5ex}[-0.5ex]{\textbf{98.3\%}}
		\\
		\rowcol
		\raisebox{-0.5ex}[-0.5ex]{4}	&		\raisebox{-0.5ex}[-0.5ex]{95.0\%}	&			\raisebox{-0.5ex}[-0.5ex]{93.2\%}	&	\raisebox{-0.5ex}[-0.5ex]{96.9\%}&	\raisebox{-0.5ex}[-0.5ex]{97.4\%}
		\\
		\bottomlinec
		
	\end{tabular}
    }
	\label{table:layer_size}
\end{table}
\begin{table}
    \caption{Hidden feature size impact on model. Instruction embedding length and layer size of 70 and 3 respectively.}
    \vspace{-1mm}
    \centering
    \resizebox{0.8\columnwidth}{!}{
    \begin{tabular}{ccccc}
		%
		\topline
		\headcol 
		
		\thead{\# hidden \\ features}  
		
		&
		F1
		&
		Precision
		&
		Recall
		&
		AUROC
		
		\\ \hline
		\raisebox{-0.5ex}[-0.5ex]{32} &  \raisebox{-0.5ex}[-0.5ex]{94.8\%} &  \raisebox{-0.5ex}[-0.5ex]{93.3\%} &  \raisebox{-0.5ex}[-0.5ex]{96.3\%} & \raisebox{-0.5ex}[-0.5ex]{97.7\%} \\ 
		\rowcol
		\raisebox{-0.5ex}[-0.5ex]{128}	&		\raisebox{-0.5ex}[-0.5ex]{95.1\%}	&		\raisebox{-0.5ex}[-0.5ex]{95.7\%}	&	\raisebox{-0.5ex}[-0.5ex]{94.5\%}	&	\raisebox{-0.5ex}[-0.5ex]{98.3\%} \\

		\raisebox{-0.5ex}[-0.5ex]{\textbf{512}}	&	
		\raisebox{-0.5ex}[-0.5ex]{\textbf{95.2\%}}	&
		\raisebox{-0.5ex}[-0.5ex]{\textbf{97.1\%}}	&
		\raisebox{-0.5ex}[-0.5ex]{\textbf{93.3\%}}	&
		\raisebox{-0.5ex}[-0.5ex]{\textbf{98.3\%}}
		\\
		
		\bottomlinec
		
	\end{tabular}
    }
	\label{table:hidden_feature_size}
\vspace{-5mm}
\end{table}


		
		
		

		
		

\vspace*{2pt}
\noindent \textbf{GNN Hidden Feature and Layers Size }
GNNs are known to experience performance degradation with an increase in network depth layers~\cite{GNNdegradation}. Therefore, our objective is to identify the point at which \mytool’s performance begins to degrade or stagnate. In our layer size experiment, presented in Table~\ref{table:layer_size}, we observe that \mytool’s performance peaks at a layer size of 3, with a significant decrease in performance at a layer size of 4. In the subsequent experiment, we aim to determine the optimal hidden feature size. As shown in Table~\ref{table:hidden_feature_size}, we test hidden feature sizes of 32, 128, and 512. However, increasing the hidden feature size does not significantly enhance \mytool’s performance. This phenomenon, observed in neural network research, suggests that networks may be prone to overfitting at higher dimensions. To address this, \mytool incorporates a dropout rate of 0.2 to mitigate potential overfitting.
Based on these findings, we conclude that a layer size of 3 and a hidden feature size of 512 are optimal for \mytool.


		
		

		
		

\vspace*{2pt}
\noindent \textbf{Instruction Embedding Length }
The embedding length defines the number of instructions from a basic block used to construct a code node embedding. Instructions in basic blocks exceeding this length are disregarded. Approximately 99.8\% of all basic blocks in our training data of usable binaries contain fewer than 70 instructions, and detailed statistics are provided in Appendix~\ref{sec:embeddinng-length}. Basic blocks with instructions below the embedding length are fully captured and used for embedding, while instructions beyond this limit are omitted.
Based on these observations, we select a final embedding length of 70 instructions, which produced optimal results for our model, as shown in Table~\ref{table:embedding_length_impact}. Embedding lengths of 80 and 90 showed a decline in performance. We hypothesize this is because longer sequences from large basic blocks begin to introduce noisy, irrelevant instructions that can dilute the semantic signal from the few critical instructions determining control flow, highlighting a trade-off between capturing more context and introducing noise.

\begin{table}[t]
    \caption{Instruction embedding length impact on model. Layer size and hidden feature size of 3 and 512 respectively.}
    \vspace{-1mm}
    \label{table:embedding_length_impact}
	\centering
	\fontsize{9pt}{9pt}\selectfont
    \resizebox{0.8\columnwidth}{!}{
	\begin{threeparttable}
		\begin{tabular}{cccccc}
			
			\topline
			\headcol 
			\thead{Embedding\\ Length}  
		
		&
		50
		&
		60
		&
		\textbf{70}
		&
		80
		&
		90
		\\ \hline

		\raisebox{-0.5ex}[-0.5ex]{AUROC}	&		\raisebox{-0.5ex}[-0.5ex]{97.4\%}	&		\raisebox{-0.5ex}[-0.5ex]{96.8\%}	&	\raisebox{-0.5ex}[-0.5ex]{\textbf{98.3\%}}	&	\raisebox{-0.5ex}[-0.5ex]{97.4\%}&	\raisebox{-0.5ex}[-0.5ex]{97.4\%} \\
		
		\rowcol
		\raisebox{-0.5ex}[-0.5ex]{F1}	&		\raisebox{-0.5ex}[-0.5ex]{94.0\%}	&		\raisebox{-0.5ex}[-0.5ex]{94.3\%}	&	\raisebox{-0.5ex}[-0.5ex]{\textbf{95.2\%}}	&	\raisebox{-0.5ex}[-0.5ex]{93.4\%}&	\raisebox{-0.5ex}[-0.5ex]{93.5\%}
		\\
		\bottomlinec
        \
			
		\end{tabular}
		
	\end{threeparttable}
 }
\vspace{-5mm}	
\end{table}

\begin{table*}[t]
	\centering
	\caption{Ablation studies on model performance. Our final F1 score is derived from Setting 10.} 
	\label{table:ablation}
	\vspace{-1mm}
	\resizebox{0.95\textwidth}{!}{
		\begin{threeparttable}
			\begin{tabular}{c | c c ccccc | c c c c}
				\topline
				\headcol
				\headcol
				
				\multirow{2}{*}{}&
				\multirow{2}{*}{}&
				\multirow{2}{*}{}&
				\multirow{2}{*}{}&
				\multirow{2}{*}{}&
				\multirow{2}{*}{}&
				\multirow{2}{*}{}&
				\multirow{2}{*}{}&
				\multicolumn{4}{c}{\textbf{Evaluation Metrics}}\\
				
				\headcol
				%
				\raisebox{1.1ex}[1.1ex]{\textbf{Setting}}&
				\raisebox{1.2ex}[1.2ex]{\thead{\textbf{Reverse} \\ \textbf{Edges}}}&
				\raisebox{1.2ex}[1.2ex]{\thead{\textbf{Data} \\ \textbf{Nodes}}}&
				\raisebox{1.2ex}[1.2ex]{\thead{\textbf{Ref-Data} \\ \textbf{Edges}}}&
				\raisebox{1.2ex}[1.2ex]{\thead{\textbf{Ref-Code} \\ \textbf{Edges}}}&
				\raisebox{1.2ex}[1.2ex]{\thead{\textbf{Function} \\ \textbf{Nodes}}}&
				\raisebox{1.2ex}[1.2ex]{\thead{\textbf{Call} \\ \textbf{Edges}}}&
				
				\raisebox{1.2ex}[1.2ex]{\thead{\textbf{Position} \\ \textbf{Encoding}}}&
				
				F1&Precision&Recall&AUROC\\
				\cline{1-12}
				\bottomlinec 
				\raisebox{-0.3ex}[-0.3ex]{1} & \raisebox{-0.3ex}[-0.3ex]{-} & \raisebox{-0.3ex}[-0.3ex]{-} & \raisebox{-0.3ex}[-0.3ex]{-} & \raisebox{-0.3ex}[-0.3ex]{-} & \raisebox{-0.3ex}[-0.3ex]{-} & \raisebox{-0.3ex}[-0.3ex]{-} & \raisebox{-0.3ex}[-0.3ex]{-} & \raisebox{-0.3ex}[-0.3ex]{83.7\%} & \raisebox{-0.3ex}[-0.3ex]{74.8\%} & \raisebox{-0.3ex}[-0.3ex]{95\%} & \raisebox{-0.3ex}[-0.3ex]{83.7\%} \\
				
				\rowcol
				\raisebox{-0.3ex}[-0.3ex]{2} & \raisebox{-0.3ex}[-0.3ex]{\grcheck} & \raisebox{-0.3ex}[-0.3ex]{-} & \raisebox{-0.3ex}[-0.3ex]{-} & \raisebox{-0.3ex}[-0.3ex]{-} & \raisebox{-0.3ex}[-0.3ex]{-} & \raisebox{-0.3ex}[-0.3ex]{-} & \raisebox{-0.3ex}[-0.3ex]{-} & \raisebox{-0.3ex}[-0.3ex]{91.5\%} & \raisebox{-0.3ex}[-0.3ex]{88.2\%} & \raisebox{-0.3ex}[-0.3ex]{95\%} & \raisebox{-0.3ex}[-0.3ex]{94.2\%} \\
				\raisebox{-0.3ex}[-0.3ex]{3} & \raisebox{-0.3ex}[-0.3ex]{-} & \raisebox{-0.3ex}[-0.3ex]{\grcheck} & \raisebox{-0.3ex}[-0.3ex]{\grcheck} & \raisebox{-0.3ex}[-0.3ex]{-} & \raisebox{-0.3ex}[-0.3ex]{-} & \raisebox{-0.3ex}[-0.3ex]{-} & \raisebox{-0.3ex}[-0.3ex]{-} & \raisebox{-0.3ex}[-0.3ex]{85.1\%} & \raisebox{-0.3ex}[-0.3ex]{76\%} & \raisebox{-0.3ex}[-0.3ex]{96.8\%} & \raisebox{-0.3ex}[-0.3ex]{85.5\%} \\
				\rowcol
				\raisebox{-0.3ex}[-0.3ex]{4} & \raisebox{-0.3ex}[-0.3ex]{-} & \raisebox{-0.3ex}[-0.3ex]{\grcheck} & \raisebox{-0.3ex}[-0.3ex]{\grcheck} & \raisebox{-0.3ex}[-0.3ex]{\grcheck} & \raisebox{-0.3ex}[-0.3ex]{-} & \raisebox{-0.3ex}[-0.3ex]{-} & \raisebox{-0.3ex}[-0.3ex]{-} & \raisebox{-0.3ex}[-0.3ex]{85.4\%} & \raisebox{-0.3ex}[-0.3ex]{75.7\%} & \raisebox{-0.3ex}[-0.3ex]{97.8\%} & \raisebox{-0.3ex}[-0.3ex]{82.7\%} \\
				
				\raisebox{-0.3ex}[-0.3ex]{5} & \raisebox{-0.3ex}[-0.3ex]{-} & \raisebox{-0.3ex}[-0.3ex]{-} & \raisebox{-0.3ex}[-0.3ex]{-} & \raisebox{-0.3ex}[-0.3ex]{\grcheck} & \raisebox{-0.3ex}[-0.3ex]{-} & \raisebox{-0.3ex}[-0.3ex]{-} & \raisebox{-0.3ex}[-0.3ex]{-} & \raisebox{-0.3ex}[-0.3ex]{85.2\%} & \raisebox{-0.3ex}[-0.3ex]{76.5\%} & \raisebox{-0.3ex}[-0.3ex]{96.2\%} & \raisebox{-0.3ex}[-0.3ex]{83.7\%} \\
				\rowcol
				\raisebox{-0.3ex}[-0.3ex]{6} & \raisebox{-0.3ex}[-0.3ex]{-} & \raisebox{-0.3ex}[-0.3ex]{-} & \raisebox{-0.3ex}[-0.3ex]{-} & \raisebox{-0.3ex}[-0.3ex]{-} & \raisebox{-0.3ex}[-0.3ex]{\grcheck} & \raisebox{-0.3ex}[-0.3ex]{-} & \raisebox{-0.3ex}[-0.3ex]{-} & \raisebox{-0.3ex}[-0.3ex]{84.3\%} & \raisebox{-0.3ex}[-0.3ex]{75.2\%} & \raisebox{-0.3ex}[-0.3ex]{95.8\%} & \raisebox{-0.3ex}[-0.3ex]{82.3\%} \\
				
				\raisebox{-0.3ex}[-0.3ex]{7} & \raisebox{-0.3ex}[-0.3ex]{-} & \raisebox{-0.3ex}[-0.3ex]{-} & \raisebox{-0.3ex}[-0.3ex]{-} & \raisebox{-0.3ex}[-0.3ex]{-} & \raisebox{-0.3ex}[-0.3ex]{-} & \raisebox{-0.3ex}[-0.3ex]{\grcheck} & \raisebox{-0.3ex}[-0.3ex]{-} & \raisebox{-0.3ex}[-0.3ex]{84.6\%} & \raisebox{-0.3ex}[-0.3ex]{75.6\%} & \raisebox{-0.3ex}[-0.3ex]{96\%} & \raisebox{-0.3ex}[-0.3ex]{82.1\%} \\
				
				\rowcol
				\raisebox{-0.3ex}[-0.3ex]{8} & \raisebox{-0.3ex}[-0.3ex]{-} & \raisebox{-0.3ex}[-0.3ex]{-} & \raisebox{-0.3ex}[-0.3ex]{-} & \raisebox{-0.3ex}[-0.3ex]{-} & \raisebox{-0.3ex}[-0.3ex]{-} & \raisebox{-0.3ex}[-0.3ex]{-} & \raisebox{-0.3ex}[-0.3ex]{\grcheck} & \raisebox{-0.3ex}[-0.3ex]{85.1\%} & \raisebox{-0.3ex}[-0.3ex]{78.8\%} & \raisebox{-0.3ex}[-0.3ex]{92.5\%} & \raisebox{-0.3ex}[-0.3ex]{86.2\%} \\
				
				\raisebox{-0.3ex}[-0.3ex]{9} & \raisebox{-0.3ex}[-0.3ex]{\grcheck} & \raisebox{-0.3ex}[-0.3ex]{\grcheck} & \raisebox{-0.3ex}[-0.3ex]{\grcheck} & \raisebox{-0.3ex}[-0.3ex]{\grcheck} & \raisebox{-0.3ex}[-0.3ex]{\grcheck} & \raisebox{-0.3ex}[-0.3ex]{\grcheck} & \raisebox{-0.3ex}[-0.3ex]{-} & \raisebox{-0.3ex}[-0.3ex]{93.7\%} & \raisebox{-0.3ex}[-0.3ex]{91.9\%} & \raisebox{-0.3ex}[-0.3ex]{95.6\%} & \raisebox{-0.3ex}[-0.3ex]{96.3\%} \\
				\rowcol
				\raisebox{-0.3ex}[-0.3ex]{\textbf{10}} & \raisebox{-0.3ex}[-0.3ex]{\grcheck} & \raisebox{-0.3ex}[-0.3ex]{\grcheck} & \raisebox{-0.3ex}[-0.3ex]{\grcheck} & \raisebox{-0.3ex}[-0.3ex]{\grcheck} & \raisebox{-0.3ex}[-0.3ex]{\grcheck} & \raisebox{-0.3ex}[-0.3ex]{\grcheck} & \raisebox{-0.3ex}[-0.3ex]{\grcheck} & \raisebox{-0.3ex}[-0.3ex]{\textbf{95.2\%}} & \raisebox{-0.3ex}[-0.3ex]{\textbf{97.1\%}} & \raisebox{-0.3ex}[-0.3ex]{\textbf{93.3\%}} & \raisebox{-0.3ex}[-0.3ex]{\textbf{98.3\%}} \\
				\bottomlinec

			\end{tabular}
		\end{threeparttable}
		
	}
\vspace{-3mm}	
\end{table*}

\subsection{Performance and Ablation Study}

Training the \mytool model took approximately 42 hours on our dataset, with an average inference time of 3 seconds per binary (with the largest binary in our dataset taking up to 117 seconds).
To evaluate the efficacy of our proposed methodology, we enable distinct features and components of the model to analyze and understand their influence on its overall performance. 
By comparing our model's performance across different settings, we can determine the relative importance of each feature and how it contributes to the overall effectiveness of our proposed approach. 
Specifically, we investigate the effects of the following features: reverse edges, data nodes, function nodes, cross-reference edges, call edges, and position encoding. 
We identified ten settings that are necessary to effectively evaluate each feature and component. 
The settings and their respective evaluation metrics are provided in Table~\ref{table:ablation}. 
Furthermore, the frequency of each new GNN feature is provided in Table~\ref{table:feature_frequency} to better judge their relative impact on the amount of information provided to the GNN.

Setting 1 serves as the baseline where no new features or components are enabled. 
In Settings 2 and 5-8, only one feature is activated in each case. Settings 3 and 4 evaluate the combination of data nodes and cross-reference edge features. 
Setting 9 tests our combined graph feature set, excluding Laplacian PE, to measure the overall impact of incorporating xRefs into \mytool's model. 
Based on Setting 9, Setting 10 adds position encoding to assess its additional impact on our model. 
Our key observations regarding the performance of the proposed model are as follows.

\noindent \textbf{Baseline } Without our proposed features, our baseline utilizes a standard CFG with a GNN to predict the indirect call target (Setting 1) yielding an F1 score of 83.7\%.

\vspace*{2pt}
\noindent \textbf{Reverse Edges }
The inclusion of reverse edges determines whether we generate bidirectional edges in our graph. 
Since CFGs and xRefs are directed, information propagation is restricted to one direction. 
By enabling reverse edges, we allow independent learning of backward execution traces alongside forward execution traces. 
Setting 2 demonstrates that incorporating this feature significantly enhances model performance, as it allows learning from backward execution traces, thereby enriching the information available to the model.

\vspace*{2pt}
\noindent \textbf{Data Node \& Reference-Data/Code Edges } 
The ``Data Node'' feature determines whether we add the data node into the ACFG. 
The ``Reference-Data Edges'' (c2d and d2d) and ``Reference-Code Edges'' (c2c and d2c)  features correspond to adding xRef edges to link the new data node to the CFG.
Overall, approximately 831K data nodes are added while 3.6M, 86K, 29, and 4 xRef edges added (for c2d, c2c, d2d, and d2c respectively). 
Despite the significantly lower frequency of d2d and d2c xRef edges—since data typically does not reference code or other data, our analysis indicates that Setting 3 (comprised of data nodes, c2d, and d2d edges) and Setting 5 (comprised of c2c and d2c edges) offers some improvements to our model. 
\mytool's performance boost with Setting 3 compared to the baseline is likely due to the data nodes being effectively connected with related, and more frequent, c2d xRef edges. A more thorough discussion on the implications of the low occurrence rate of d2d and d2c xRef edges can be found in $\S$\ref{sec:data_nodes_limitations}.
\mytool with Setting 5 sees a modest performance boost as including xRefs to specific code nodes captures additional relationships that are not directly provided in traditional CFGs.
Introducing these xRef edges provides CFGs richer semantic context, particularly in scenarios where indirect references to code nodes influence program behavior. This allows the model to identify code segments that operate on the same data, revealing shared behaviors or patterns, which is valuable in distinguishing different tasks with similar structure. 
Setting 4 shows that the combination of all three features yields the best performance. 
Therefore, we conclude that adding xRef information is crucial to \mytool's success.


\begin{table} 
    \caption{Feature frequency statistics.}
     \vspace{-1mm}
	\label{table:feature_frequency}
    
	\centering
	\fontsize{9pt}{9pt}\selectfont
    \resizebox{0.8\columnwidth}{!}{
	\begin{threeparttable}
		\begin{tabular}{ccc}
			
			\topline
			\headcol 
			
			\textbf{Feature Type}
			
            &
			\textbf{Feature Name} 
			&
			\textbf{Count}

			\\ \hline
			\raisebox{-0.5ex}[-0.5ex]{Node} & \raisebox{-0.5ex}[-0.5ex]{Code} & \raisebox{-0.5ex}[-0.5ex]{5.2M} \\ 
			\rowcol
            
			\raisebox{-0.5ex}[-0.5ex]{Node} & \raisebox{-0.5ex}[-0.5ex]{Data} & \raisebox{-0.5ex}[-0.5ex]{831K} \\ 

			\raisebox{-0.5ex}[-0.5ex]{Node} & \raisebox{-0.5ex}[-0.5ex]{Function} & \raisebox{-0.5ex}[-0.5ex]{350K} \\
   
        \bottomlinec
        \rowcol
        \raisebox{-0.5ex}[-0.5ex]{Edge} & \raisebox{-0.5ex}[-0.5ex]{Control Flow} & \raisebox{-0.5ex}[-0.5ex]{6.3M} \\ 

        \raisebox{-0.5ex}[-0.5ex]{Edge} & \raisebox{-0.5ex}[-0.5ex]{Code-to-Function} & \raisebox{-0.5ex}[-0.5ex]{5.2M} \\

        \rowcol
        \raisebox{-0.5ex}[-0.5ex]{Edge} & \raisebox{-0.5ex}[-0.5ex]{Code Call} & \raisebox{-0.5ex}[-0.5ex] {2.3M} \\ 

        \raisebox{-0.5ex}[-0.5ex]{Edge} & \raisebox{-0.5ex}[-0.5ex]{Code2Data} & \raisebox{-0.5ex}[-0.5ex] {3.6M} \\ 

        \rowcol
        \raisebox{-0.5ex}[-0.5ex]{Edge} & \raisebox{-0.5ex}[-0.5ex]{Data2Data\tnote{a}} & \raisebox{-0.5ex}[-0.5ex]{29}  \\

        \raisebox{-0.5ex}[-0.5ex]{Edge} & \raisebox{-0.5ex}[-0.5ex]{Code2Code\tnote{b}} & \raisebox{-0.5ex}[-0.5ex]{86K} \\ 

        \rowcol
        \raisebox{-0.5ex}[-0.5ex]{Edge} & \raisebox{-0.5ex}[-0.5ex]{Data2Code\tnote{a}} & \raisebox{-0.5ex}[-0.5ex]{4\tnote{c}} \\

		\bottomlinec
			
		\end{tabular}
		\begin{tablenotes}
			\footnotesize
			\item[a] Most data does not reference other code or data; instead, it is typically referenced by code.
            \item[b] Distinct from standard CFG control flow edges.
            \item[c] The count for Data2Code edges is low as our symbolization process primarily identifies direct references to function entry points within the data section. It may not resolve all jump table structures, which are a known challenge in binary analysis and a subject for future work.
		\end{tablenotes}
		
	\end{threeparttable}
    }
\vspace{-5mm}	
\end{table}

\vspace*{2pt}
\noindent \textbf{Function Node }
Neural networks often struggle with graphs containing distant nodes, as they are unable to efficiently transfer information across these nodes~\cite{GNNdegradation,sanchez-lengeling2021a}. 
To mitigate this issue, we incorporate an aggregate node~\cite{aggregateNode2017,aggregateNode2_2018}, referred to as the ``Function Node'' (Setting 6) in our context, aiming to enhance performance. Approximately 350K function nodes and the associated 5.2M code-to-function edges can be created for our model, as seen in Table~\ref{table:feature_frequency}.
This new node connects all of a function's basic block nodes with a special edge type. However, as indicated in  Table~\ref{table:ablation}, this feature does not significantly impact the model's performance.


\vspace*{2pt}
\noindent \textbf{Call Edges }
In Setting 7, we add a new edge type to distinguish direct ``call edges'' from other edges in the original CFG. Table~\ref{table:feature_frequency} shows that 2.3 million new call edges can be added to the model. Given this,
our findings indicate that including direct call edge labels offers marginal improvement in predicting icall targets.


\vspace*{2pt}
\noindent \textbf{Positional Encoding (PE) }
Setting 8 demonstrates that incorporating PE may enhance our model. PE allows the graph to locate specific nodes by assigning unique positional information to each node, 
thereby improving the model's ability to distinguish between different nodes within the graph. After adding PE, \mytool has an F1 score of 85.1\%.

\vspace*{2pt}
\noindent \textbf{All without PE }
To demonstrate the effectiveness of our ACFG in improving the icall resolution, we combined all of our proposed graph features (Setting 9), 
yielding an F1 score of 93.7\%, which already outperforms the optimized Callee.  


\vspace*{2pt}
\noindent \textbf{Overall Performance: All with PE }
To further improve upon the results of Setting 9, we incorporated Laplacian PE as shown in Setting 10. This adjustment had a beneficial impact on the model, 
notably increasing precision and leading to the highest F1 and AUROC scores observed.

\vspace*{2pt}
\noindent \textbf{Impact of Position Encoding on Precision vs. Recall } 
Adding position encoding (Setting 10) makes \mytool more conservative. Because PE helps the model learn the graph's structure, it becomes more confident about predictions that fit typical patterns, boosting precision. However, this same focus on structure means the model can sometimes miss true positives that have unusual or rare control flow paths, causing a slight drop in recall (from 95.6\% to 93.3\%).
Specifically, we observed that icall sites in deeply nested call chains (rare in the training data) were occasionally misclassified when position encoding overemphasized graph distance. 
While high precision is often preferred in downstream security applications (e.g., binary-level CFI, where false positives can disrupt legitimate control flows), future work could investigate hybrid positional encoding strategies (e.g., combining Laplacian encoding with learned attention mechanisms) to mitigate this trade-off and recoup recall.

\subsection{Comparative Study } 
\label{sec:CalleeComparison}
We focus our primary comparison against Callee~\cite{Callee}, as the original Callee paper demonstrated its superiority over prior works such as TypeArmor~\cite{TypeArmor} and BPA~\cite{BPA}. Therefore, we directly compare \mytool with Callee, representing the current state-of-the-art, and also evaluate the efficacy of AttnCall~\cite{AttnCall2024} by simulating its direct-call learning methodology. 
Callee's released pre-trained model performs poorly in our testing environment, achieving an F1 score of 43.9\%. However, the current publicly available materials for Callee does not include training code which prevents us from conducting additional training, input fine-tuning, and further optimizations.\footnote{Callee's authors unfortunately did not respond to our multiple requests regarding reproducing their work or providing their training materials. Furthermore, their released dataset only contains the preprocessed data where cross-reference information utilized by \mytool has already been lost.}
By re-implementing Callee's missing materials based on its paper description, we were able to train Callee on our comprehensive dataset, achieving a significantly improved F1 score of 89.9\%. This re-implemented version serves as our main baseline for comparison.

\begin{figure} [t]
	\centering
	\includegraphics[width=0.9\columnwidth]{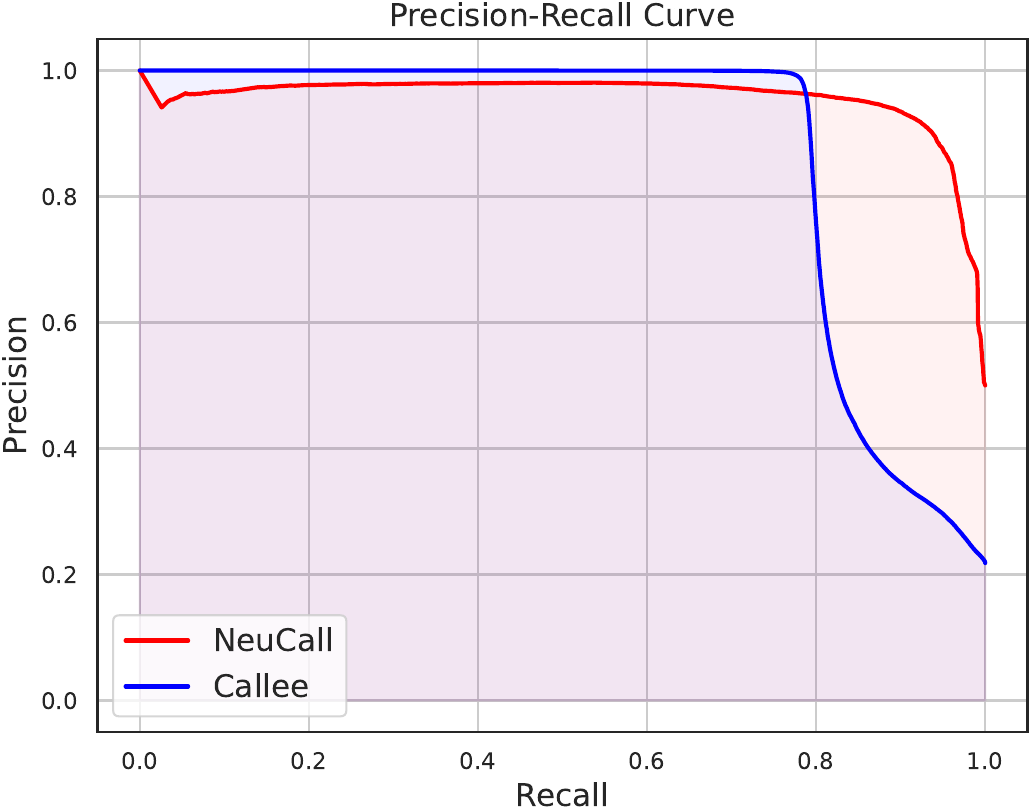}
	
	\vspace{-1mm}
	\caption{Precision-recall curves for \mytool and Callee}
	\label{fig:precision-recall}
	\vspace{-5mm}
\end{figure}

\vspace*{2pt}
\noindent \textbf{Callee }
Several factors may contribute to the initial poor performance of the pretrained version of Callee, though no single issue can be identified conclusively as we do not have access to the original training code. Overall, we hypothesize that the primary factors are model overfitting and limited generalization within its dataset.

Fortunately, we were able to re-implement a training script for Callee, enabling us to fine-tune and train it on our dataset for a fair comparison with \mytool. To optimize Callee’s performance, we applied the same balanced sampling process used to test our model, ensuring an equal representation of valid targets and invalid targets. Addressing imbalanced datasets continues to be a significant research challenge in the optimization of ML and neural network tasks~\cite{deZarza2023OptimizingNN, HUANG2022NN_for_imbalanced_data}.

After applying our optimizations, Callee's F1 score rises to 89.9\%. 
In comparison, \mytool achieves an F1 score of 95.2\% based on Setting 10, demonstrating superior performance. 
We plot the precision-recall curve for both tools in Figure~\ref{fig:precision-recall}, which shows that \mytool outperforms Callee with significantly less trade-off for higher recall, 
indicating that \mytool is more robust.
The discrepancy between Callee's reported and evaluated performance can be attributed to the fact that we cannot verify their released dataset. 
Specifically, the authors only provide their preprocessed dataset without the associated binaries from which the data were derived.



\vspace*{2pt}
\noindent \textbf{AttnCall }
Although we are unable to conduct a comprehensive evaluation of AttnCall due to its non-reproducibility, we adopt their direct-call (dcall) training methodology~\cite{AttnCall2024} to assess its effectiveness. Specifically, we train \mytool exclusively on dcalls to predict icall targets. This methodology predictably underperforms, yielding an F1 score of 33.3\% for icall target prediction. Consequently, their assumption that dcalls can replace icalls for training and testing is not valid.

\subsection{Optimization Impact \& Security Application}

\textbf{First}, we evaluated \mytool's performance on a set of Arch Linux binaries compiled with varying compiler optimization levels (O0, O1, O2, O3). The results, detailed in Appendix~\ref{sec:Aopt}, show that \mytool maintains relatively consistent F1 scores across these different levels, indicating resilience to common code transformations introduced by optimizations. Notably, performance peaked for O1-optimized binaries and saw only a slight decline at the O3 level, suggesting that \mytool's graph-based representation effectively captures structural information pertinent to indirect calls even when faced with significant code modifications. This robustness is crucial for real-world applications where binary optimization levels are often diverse or unknown.
\textbf{Second}, we explored the applicability of \mytool's predictions for downstream security tasks, specifically binary-level Control Flow Integrity (CFI). As presented in Appendix~\ref{sec:Acfi}, we compared the granularity of the indirect call target sets predicted by \mytool against those from Callee and the source-level baseline LLVM-CFI, using the Average Indirect Call Target (AICT) metric. The experiment demonstrates that \mytool achieves significantly more precise target sets (lower AICT) than Callee, substantially closing the gap towards the precision offered by compiler-level CFI approaches. This suggests \mytool can provide a more effective foundation for implementing practical binary-only CFI enforcement compared to prior ML-based techniques.

\section{Discussion}
\label{sec:discussion}


Despite \mytool's strong performance, several limitations and opportunities for future work remain.

\vspace*{2pt}
\noindent \textbf{Ground Truth Dilemma }\label{sec:gt}
Establishing accurate ground truth for indirect call resolution is inherently challenging, balancing the trade-offs between soundness and completeness. Dynamic analysis, while sound in recording actual execution targets, often lacks completeness due to path coverage limitations. Conversely, static analysis leveraging source code offers better completeness by using type information but may suffer from soundness issues, potentially yielding false positives. For generating our training and evaluation dataset, we opted for a compiler-level analysis approach, which provides scalability and leverages source-level information when available ~\cite{Lin2023TypeSqueezer,lu2019does}. Specifically, we utilized TyPro~\cite{bauer2022typro} due to its relatively low false positive rate and robust engineering quality, enabling the collection of a sufficiently large dataset. We acknowledge that this choice means our current dataset inherits TyPro's intrinsic limitations. However, it is crucial to emphasize that \mytool itself is designed modularly and is not tied to any specific ground truth generation tool like TyPro. \mytool operates on the generated dataset, meaning its core architecture can readily leverage improved datasets. As more advanced compiler-level type analysis techniques emerge~\cite{ xia2024deeptype, liu2024improving, cai2024unleashing}, they can produce datasets with higher precision and recall. Such improved datasets can be seamlessly integrated into our pipeline to train more accurate \mytool models, highlighting the adaptability of our approach to advancements in prerequisite static analysis tools.

\vspace*{2pt}
\noindent \textbf{Role and Limitation of Data Nodes in Augmented CFGs }
\label{sec:data_nodes_limitations}
The inclusion of data nodes in our augmented CFGs enhances the representation of code-data relationships, providing a more comprehensive view of control flow. These nodes serve as bridges, connecting code nodes that reference the same data points and capturing indirect relationships that traditional CFGs may overlook. This bridging capability is particularly valuable in cases where traditional CFGs alone are insufficient to fully characterize program behavior.
A noteworthy observation in our implementation is that the number of xRef edges originating from data nodes is relatively low. This is not a limitation of our approach but rather reflects the nature of binary programs, where most data lacks explicit xRefs. Much of the data accessed in binaries is localized or lacks external references that would produce outgoing edges. Nevertheless, data nodes are crucial for linking code segments with shared data access points, even in the absence of direct xRefs.
Another consideration is the challenge of determining data structure boundaries. In our current model, data is treated as fixed-size units, simplifying the representation but potentially missing more complex relationships within structured or multi-field data. While advanced program analysis techniques could infer data structure boundaries~\cite{Zhiqiang10,Howard11} and uncover additional implicit xRef edges, these methods are not perfect and may introduce false positives. Balancing the trade-off between capturing richer data relationships and maintaining precision is a key consideration for future enhancements.

\vspace*{2pt}
\noindent \textbf{Indirect Jumps }
Another form of indirect control flow is indirect jumps, which typically arise from switch-case structures in source code~\cite{Federico16, JTR},
a function's return instruction, and the effects of tail-call optimization~\cite{Meng16}.
Our current model is designed specifically to recover icalls and does not address the recognition of indirect jump targets due to the lack of reliable training data. 
Many existing CFI approaches offer insufficient protection against indirect jumps, often providing only coarse-grained safeguards, as noted by Burow et al.~\cite{burow2017control}. 
Should the issue of obtaining a high-quality training set be resolved, \mytool could be extended to predict indirect jump targets.



\vspace*{2pt}
\noindent \textbf{Graph Positional Encoding for Heterogeneous GNNs }
Currently, there is no standardized or well-defined methodology for positional encoding (PE) in heterogeneous graphs. 
Zhao et al.~\cite{ZHAO2022} also highlights this issue, noting their use of Laplacian PE as a substitute. 
Despite being designed for homogeneous graphs, our ablation study results in Table~\ref{table:ablation} demonstrates that PE has a significant beneficial impact on our model. 
While developing a robust PE method for heterogeneous graphs is beyond the scope of this paper, we encourage future research to address this gap, as it could potentially further enhance our model's performance.



\section{Conclusion}\label{sec:conclusion2}



This paper presents \mytool, a novel framework for resolving indirect calls in stripped binaries using GNNs. \mytool advances the state of the art through two key innovations: a new xRef-augmented interprocedural CFG (ICFG) that incorporates both code and data cross-references to capture richer program semantics, and a compiler-guided strategy for collecting high-quality training data that accurately reflects real-world calling behavior. Our relational graph convolutional model further enhances prediction accuracy by leveraging structural and semantic relations encoded in the augmented ICFG.
Comprehensive evaluations on real-world binaries demonstrate that \mytool significantly outperforms state-of-the-art approaches in both accuracy and robustness. 
We believe \mytool represents a step forward in applying neural graph learning to binary program understanding and opens new avenues for advancing analysis of low-level software.

\newpage
\section*{Ethics Considerations}

Our research focuses on improving indirect call target prediction in binary code analysis using GNNs. This work aims to advance the accuracy and reliability of static analysis tools, thereby enhancing software security practices and protecting users from vulnerabilities that attackers could exploit through indirect control flows.

Firstly, the datasets utilized for training and evaluating our model were compiled from publicly accessible sources, including GitHub and the Arch User Repository. The collected binaries do not contain sensitive or private user information, and our dataset generation methods adhere strictly to ethical standards, ensuring no infringement on individual or organizational privacy rights.

Secondly, our research methodologies involve no direct interaction with live user environments, real-time networks, or operational systems. Instead, evaluations are conducted exclusively within controlled laboratory settings. This approach guarantees that our research activities do not inadvertently cause economic harm, disrupt services, or negatively affect users' psychological well-being.

Lastly, we commit to transparency and openness by releasing our prototype implementation, pretrained models, and curated datasets through publicly available platforms such as Zenodo. This practice facilitates reproducibility, verification, and further development within the broader research community, thereby promoting ethical collaboration and progress in binary code analysis.

\bibliographystyle{unsrt} 
\bibliography{path-exploration,binary-code,code-reuse,debloating,mips,other,jiang,others,gnn,binary,asplos}




\appendices

\setcounter{table}{0}
\setcounter{figure}{0}
\renewcommand{\thetable}{A\arabic{table}}
\renewcommand{\thefigure}{A\arabic{figure}}


\section{Evaluating BPE Tokenization for Numeric Address Embedding}
\label{sec:BPE}
\subsection{Motivation}
Subword tokenization techniques, such as Byte Pair Encoding (BPE), offer a potential way to embed numeric addresses and mitigate the Out-of-Vocabulary (OOV) issues frequently encountered in natural language processing~\cite{bostrom-durrett-2020-byte}. While standard tokenization in binary analysis often replaces addresses with generic tokens (e.g., [addr]), losing vital cross-reference information, BPE offers a way to represent numeric values without discarding them entirely. However, our core argument in this paper is that numeric addresses in binary code represent relational information (cross-references) that are best captured structurally, for instance, as edges in an augmented Control Flow Graph (ACFG) like \mytool employs, rather than as sequential tokens. We hypothesize that simply applying BPE to addresses, while potentially handling OOV, fails to capture their semantic meaning and may even introduce noise, degrading model performance.

\subsection{Experimental Setup}
To investigate the efficacy of BPE for numeric address embedding, we conducted an experiment using the Callee model  as a baseline. Callee, discussed in our related work (Section 2.2 ), utilizes a doc2vec model which, like many NLP-inspired approaches, faces challenges with representing numeric addresses effectively.

We modified the Callee framework to specifically apply BPE tokenization only to the numeric address values encountered during its preprocessing phase. The rest of the Callee model and its training procedure remained unchanged. We tested BPE with varying vocabulary sizes, controlled by the number of byte pairs allowed: 8, 32, and 512. We also included the performance of the original Callee model (equivalent to using 0 byte pairs for numeric addresses, relying on its default handling) as a baseline for comparison. The performance was measured using the F1 score on the indirect call target prediction task.

\begin{table} [ht]
	\centering
	\caption{The F1 scores for the Callee model with different BPE settings applied to numeric addresses.} 
	\label{table:bpetable}
	\fontsize{7pt}{7pt}\selectfont
    \resizebox{0.7\columnwidth}{!}{
	\begin{threeparttable}
		\begin{tabular}{lc}
			
			\topline
			\headcol \raisebox{-0.2ex}[-0.2ex]{Setting} & \raisebox{-0.2ex}[-0.2ex]{F1 Score} \\
            \midline
			\raisebox{-0.2ex}[-0.2ex]{No BPE (Original Callee)} & \raisebox{-0.2ex}[-0.2ex]{88.9\%} \\
			\rowcol
			\raisebox{-0.2ex}[-0.2ex]{BPE (8 byte pairs)} & \raisebox{-0.2ex}[-0.2ex]{84.3\%} \\
			\raisebox{-0.2ex}[-0.2ex]{BPE (32 byte pairs)} & \raisebox{-0.2ex}[-0.2ex]{79.6\%} \\
			\rowcol
			\raisebox{-0.2ex}[-0.2ex]{BPE (512 byte pairs)} & \raisebox{-0.2ex}[-0.2ex]{79.0\%} \\
			\bottomlinec
		\end{tabular}
		\end{threeparttable}
        }
\vspace{-4mm}
\end{table}

\subsection{Results and Discussion}
The results in Table~\ref{table:bpetable} clearly demonstrate that applying BPE tokenization to numeric addresses negatively impacts the performance of the Callee model. As the complexity of the BPE tokenization increases (i.e., more byte pairs allowed, potentially creating finer-grained subword units for addresses), the F1 score drops significantly compared to the baseline where addresses were handled by Callee's original, simpler tokenization scheme. This supports our hypothesis that BPE, despite its ability to handle OOV numbers, is not suitable for embedding numeric addresses in this context. 

Our concern lies in the semantic meaninglessness of the resulting embeddings. Even if BPE avoids OOV problems by breaking addresses into subword tokens (e.g., splitting an address into constituent numeric parts or common prefixes), these tokens lack consistent semantic meaning across different binaries or even within the same binary. An address like 0x1234ABCD might be tokenized differently based on the BPE vocabulary learned from the corpus, and its resulting embedding carries no inherent relational information. The core issue is that subword tokenization cannot resolve the fundamental arbitrariness of addresses – an address value like 0x401000 in one binary is semantically unrelated to 0x401000 in a different binary compiled independently.

Furthermore, attempting to embed these arbitrary numeric sequences can introduce noise, potentially diluting the meaningful signals derived from other tokens like opcodes and instruction mnemonics, which have more consistent semantics. While we acknowledge that BPE can theoretically mitigate OOV issues for numbers, its application to addresses fails to provide meaningful semantic representations and, as shown by the results, degrades predictive accuracy.

This contrasts sharply with \mytool's approach, which treats addresses not as tokens to be embedded in a sequence, but as pointers representing relationships (cross-references). By explicitly modeling these relationships as edges within an augmented graph structure (ACFG), \mytool preserves the crucial semantic role of addresses in defining program structure and control flow, leading to superior performance in resolving indirect calls. This experiment reinforces our position that structural graph-based representations are more appropriate for capturing the relational nature of addresses in binary analysis than sequential tokenization methods like BPE.

\begin{figure} [t] 
	\centering
	\includegraphics[width=0.9\columnwidth]{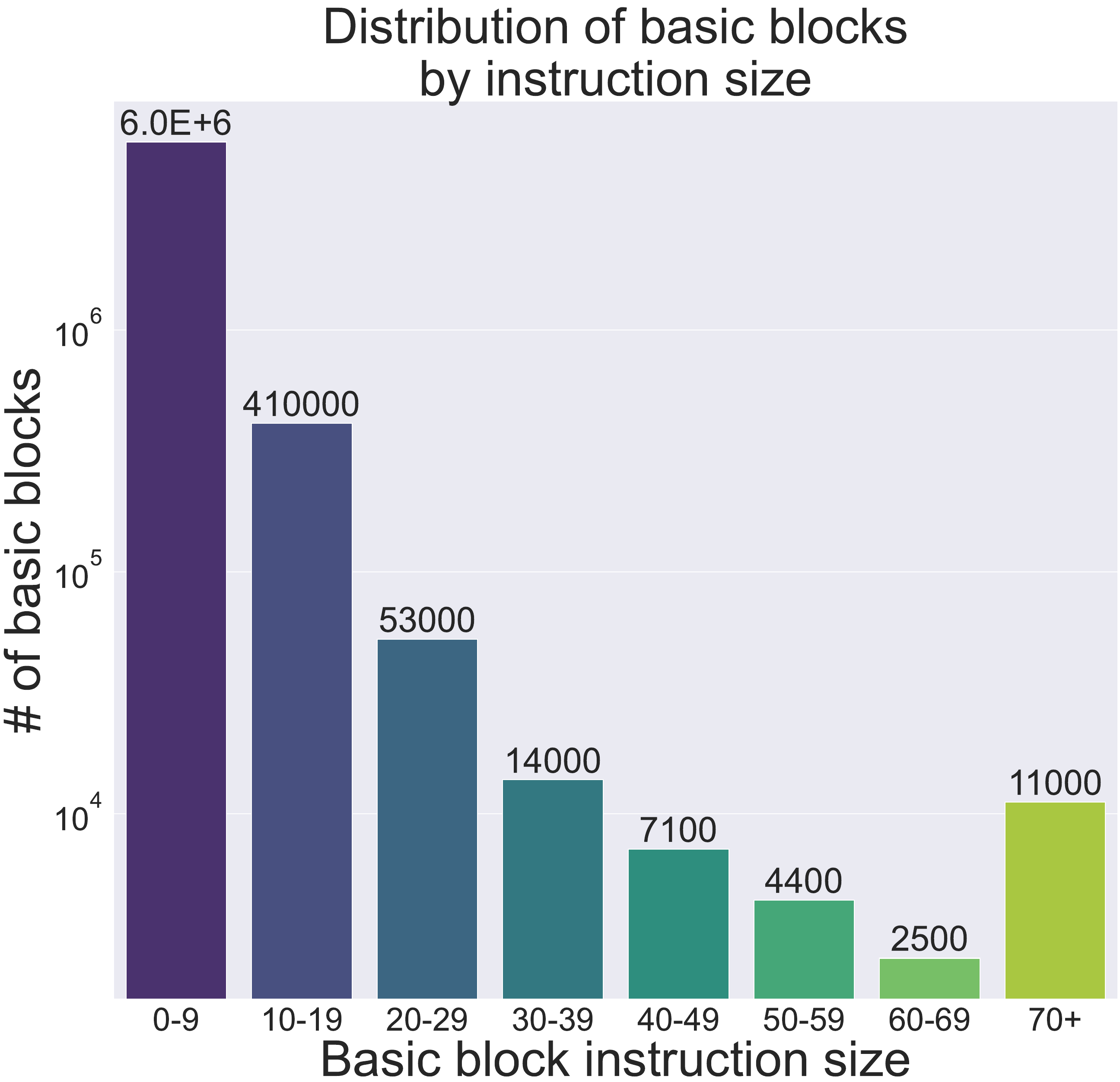} 
	\caption{Approximate number of basic blocks per instruction length. After compiling source code from our dataset (Arch User Repository~\cite{Arch} and GitHub Repositories), we have a total of $6,513,030$ basic blocks. About 92.0\% of basic blocks contain less than 10 instructions, and 99.8\%  of them have less than 70 instructions.} 
	\label{fig:insperbb}
\end{figure}

\section{Instruction Embedding Length} \label{sec:embeddinng-length}

As shown in Figure~\ref{fig:insperbb}, approximately 99.8\% of all basic blocks in our training data of usable binaries contain fewer than 70 instructions. Thus, we select a final embedding length of 70 instructions.

\begin{table} [ht]
	\centering
	\caption{\mytool performance on different optimizations.} 
	\label{table:cross-opt}
	\fontsize{9.5pt}{9.5pt}\selectfont
    \resizebox{0.9\columnwidth}{!}{
	\begin{threeparttable}
		\begin{tabular}{ccccc}
			
			\topline
			\headcol 
			\raisebox{-0.2ex}[-0.2ex]{Opt. Level} & \raisebox{-0.2ex}[-0.2ex]{F1} & \raisebox{-0.2ex}[-0.2ex]{Precision} & \raisebox{-0.2ex}[-0.2ex]{Recall} & \raisebox{-0.2ex}[-0.2ex]{AUROC} \\
			\midline
			\raisebox{-0.2ex}[-0.2ex]{O0} & \raisebox{-0.2ex}[-0.2ex]{94.7\%} & \raisebox{-0.2ex}[-0.2ex]{97.4\%} & \raisebox{-0.2ex}[-0.2ex]{92.2\%} & \raisebox{-0.2ex}[-0.2ex]{98.3\%} \\
			\rowcol
			\raisebox{-0.2ex}[-0.2ex]{O1} & \raisebox{-0.2ex}[-0.2ex]{96.8\%} & \raisebox{-0.2ex}[-0.2ex]{97.0\%} & \raisebox{-0.2ex}[-0.2ex]{96.6\%} & \raisebox{-0.2ex}[-0.2ex]{97.7\%} \\
			
			\raisebox{-0.2ex}[-0.2ex]{O2} & \raisebox{-0.2ex}[-0.2ex]{94.7\%} & \raisebox{-0.2ex}[-0.2ex]{90.9\%} & \raisebox{-0.2ex}[-0.2ex]{98.9\%} & \raisebox{-0.2ex}[-0.2ex]{97.6\%} \\
			\rowcol
			\raisebox{-0.2ex}[-0.2ex]{O3} & \raisebox{-0.2ex}[-0.2ex]{93.6\%} & \raisebox{-0.2ex}[-0.2ex]{94.3\%} & \raisebox{-0.2ex}[-0.2ex]{92.9\%} & \raisebox{-0.2ex}[-0.2ex]{96.9\%} \\
			\bottomlinec
		\end{tabular}
		\end{threeparttable}
        }
\end{table}

\section{Optimization-Level Evaluation}
\label{sec:Aopt}
We test our pre-trained \mytool model on our collected Arch Linux binaries with varying optimizations to observe how optimization levels can influence the model. The evaluation metrics are presented in Table~\ref{table:cross-opt}.

\begin{table*} [t]
    \centering
    
    \caption{Binary CFI Experiment: \mytool provides a more refined AICT metric compared to Callee and thus closing the gap between LLVM-CFI. ``AT'' in Column 4 is short for Address-Taken functions.}
    \resizebox{0.7\textwidth}{!}{
    \begin{tabular}{lrrrrrr}
        \topline
        \headcol
        
        \textbf{Binary} & \textbf{\# of Functions} & \textbf{\# of iCalls} & \textbf{\# of AT}  & \thead{\textbf{Callee }\\ \textbf{AICT}} & \thead{\textbf{\mytool }\\ \textbf{AICT}} & \thead{\textbf{LLVM-CFI }\\ \textbf{AICT}} \\
        \midline
        certutil & 1240	& 3 & 408 &  {4.3} & 5.0 & 5.0 \\
        \rowcol
        pk12util & 913 & 3 & 345 &  {2.8} & 7.3 & 6.7 \\
        
        signmar & 301 & 1 & 126 &  {2.5} & 2.0 & 2.0 \\
        \rowcol
        libfreeblpriv3.so & 4393 & 177 & 1115 & 93.9 & {35.1} & 29.3 \\
        libnspr4.so & 2008 & 145 & 906 & 181.4 & {44.8} & 10.6 \\
        \rowcol
        libmozsqlite3.so & 7110 & 907 & 905 & 202.8 & {89.8} & 33.1 \\
        libnss3.so & 5682 & 383 & 2548 & 330.9 & {64.3} & 15.3 \\
        \rowcol
        libssl3.so & 3482 & 373	& 1410 & 82.2 & {26.0} & 19.5 \\
        \bottomline
    \end{tabular}
    
    \label{table:cfi}
}
\end{table*}

The results indicate a good degree of robustness for \mytool against common compiler optimizations. The relative stability in F1 scores across O0, O1, O2, and O3 suggests that the structural and semantic features captured by \mytool's Cross-References Augmented Control Flow Graph (ACFG) representation are somewhat resilient to the code transformations performed by the compiler. The peak performance observed at the O1 level might suggest that this level strikes a favorable balance for \mytool – code is potentially "cleaner" than O0 due to basic optimizations removing redundancy, but not yet subjected to the highly aggressive transformations of O3 (such as extensive function inlining or complex loop vectorizations) that might obscure some higher-level structural patterns or significantly increase graph complexity. The slight decrease in performance at O3 could be attributed to these aggressive transformations potentially making the mapping between call sites and potential target functions more convoluted or introducing patterns less represented in the original training data.

Overall, this experiment demonstrates that \mytool generalizes reasonably well to binaries compiled with different optimization flags, reinforcing its potential for practical application in analyzing real-world software where optimization levels vary or are unknown.

\section{Case Study: Target Set Granularity for Security Hardening}
\label{sec:Acfi}
To demonstrate the practical utility of \mytool in security applications, we conducted an experiment to evaluate its effectiveness in simulating a binary-level control flow integrity (CFI) policy. Using Firefox as a case study, we compiled its source code, extracted indirect call targets, and analyzed \mytool's predictions in comparison with LLVM-CFI~\cite{clang-cfi} and Callee~\cite{Callee}.

\vspace*{2pt}
\noindent \textbf{Experimental Setup}
To demonstrate the practical utility of \mytool's predictions for downstream security tasks, we conducted a case study to evaluate the granularity of its predicted indirect call target sets. While our model does not achieve the 100\% recall required for a sound CFI enforcement policy, the precision of the target set is critical for reducing the attack surface in security hardening applications. A smaller, more precise set of possible targets (a lower Average Indirect Call Target, or AICT) provides a stronger foundation for security tools by minimizing the number of valid but unintended gadgets an attacker can use.

In this case study, we compare the AICT of the target sets predicted by \mytool against those from Callee and the source-level baseline, LLVM-CFI. The goal is not to propose \mytool as a standalone CFI solution, but to quantify how effectively it closes the precision gap between binary-level analysis and compiler-level analysis, thereby providing a more robust basis for binary hardening techniques.

\vspace*{2pt}
\noindent \textbf{Results and Analysis }
Our results, presented in Table~\ref{table:cfi}, demonstrate that \mytool consistently achieves AICT values closer to LLVM-CFI compared to Callee. For executables with a large number of indirect calls, such as ``libnspr4.so'' and ``libnss3.so,'' Callee exhibited substantially higher AICT values, indicating greater over-approximation and less alignment with LLVM-CFI’s precision. Across the 8 binaries analyzed, \mytool reduced the AICT gap with LLVM-CFI by over 80\% relative to Callee, highlighting \mytool’s ability to provide a significantly more precise approximation of source-level CFI policies.

\vspace*{2pt}
\noindent \textbf{AIR and AICT } 
Average Indirect Target Reduction (AIR)~\cite{zhang2013CFI4COTS} and Average Indirect Call Targets (AICT)~\cite{lu2019does} are widely used performance metrics in the CFI domain~\cite{BPA}. 
While these metrics provide a practical measure of the granularity of predicted target sets, recent research has highlighted their limitations, particularly regarding soundness and completeness~\cite{burow2017control,frassetto2022cfinsight}. 
Specifically, AIR and AICT cannot guarantee that all icall targets have been accurately identified or that no extraneous targets are included.
In our study, we report the F1 score to evaluate correctness, capturing the precision and recall of \mytool’s predictions. AICT, on the other hand, serves a complementary purpose, enabling a comparative analysis of target set granularity between \mytool, Callee, and LLVM-CFI. While AICT does not fully address correctness, it remains a valuable metric for illustrating how closely \mytool’s target sets align with the source-level baseline provided by LLVM-CFI.

	

\end{document}